\documentclass[runningheads]{llncs}
\usepackage[utf8]{inputenc}
\usepackage{times}
\usepackage[dvipsnames]{xcolor}
\usepackage{url,mathtools,amssymb,tikz}
\DeclarePairedDelimiter{\ceil}{\lceil}{\rceil}
\allowdisplaybreaks
\usetikzlibrary{automata,arrows,calc}
\tikzset{->,>=stealth',auto,node distance=2cm,thick,initial text=}
\tikzstyle{accepting}=[path picture={%
  \draw let 
    \p1 = (path picture bounding box.east),
    \p2 = (path picture bounding box.center)
    in
    (\p2) circle (\x1 - \x2 - 2pt);
 }]
\usepackage{caption}
\usepackage{subcaption}
\definecolor{green1}{rgb}{0, 0.5, 0}
\definecolor{red1}{rgb}{0.64, 0, 0}
\usepackage{graphicx}
\usepackage{hyperref}

\usepackage[disable]{todonotes} 
\usepackage{nicefrac} 
\usepackage{bbm} 
\usepackage[numbers,sort&compress]{natbib}
\begin{document}
\title{Polarization and Belief Convergence of Agents in Strongly-Connected Influence Graphs
}
%
%
\author{M\'{a}rio S. Alvim\inst{1} \and
Bernardo Amorim\inst{1} \and
Sophia Knight\inst{2} \and
Santiago Quintero\inst{3} \and
Frank Valencia\inst{4}
}
\authorrunning{Alvim et al.}
\titlerunning{Polarization and Belief Convergence}
%
\institute{Department of Computer Science, Universidade Federal de Minas Gerais \and
Department of Computer Science, University of Minnesotta Duluth \and
LIX, \'{E}cole Polytechnique de Paris \and
CNRS-LIX, \'{E}cole Polytechnique de Paris 
}
\maketitle              
\begin{abstract}
We describe a model for polarization in multi-agent systems based on Esteban and Ray's classic measure of polarization from economics. Agents evolve by updating their beliefs (opinions) based on the beliefs of others and an underlying influence graph. We show that polarization eventually disappears (converges to zero) if the influence graph is strongly-connected. If the influence graph is a circulation we determine the unique belief value all agents converge to. For clique influence graphs we determine the time after which agents will reach a given difference of opinion. Our results imply that if polarization does not disappear then either there is a disconnected subgroup of agents or some agent influences others more than she is influenced. Finally, we show that polarization does not necessarily vanish in weakly-connected graphs, and illustrate the model with a series of case studies and simulations giving some insights about polarization.
\todo{Mario on 2020-10-19: Also emphasize on abstract that we
consider convergence of polarization even in cases consensus isn't
reached?}

\end{abstract}
%
%
%

\newcommand{\R}{\mathbb{R}} 
\newcommand{\IfunM}{\Inter_{min}} 
\newcommand{\mx}[1]{max^{#1}} 
\newcommand{\mn}[1]{min^{#1}} 
\newcommand{\IfunR}[2]{\Inter^{\textit{real}}_{#1,#2}} 
\newcommand{\Path}[2]{\infl{#1}{#2}} 
\newcommand{\CBfun}[3]{f^{#3}_{\agent{#1},\agent{#2}}} 
\newcommand{\CBfunM}{f_{min}}


\newcommand{\Agents}{\mathcal{A}} 
\newcommand{\agent}[1]{#1} 
\newcommand{\Blf}{B} 
\newcommand{\Blft}[1]{\Blf^{#1}} 
\newcommand{\Bfun}[2]{\Blf^{#2}_{\agent{#1}}} 
\newcommand{\Bapp}[3]{\Blf^{#3}_{\agent{#1}{\mid}\agent{#2}}} 


\newcommand{\qm}[1]{``#1''}

\newcommand{\cali}{\mathcal{I}}
\newcommand{\calx}{\mathcal{X}}
\newcommand{\caly}{\mathcal{Y}}
\newcommand{\calt}{\mathcal{T}}

\newcommand{\Pol}{\rho}
\newcommand{\Pfun}[1]{\Pol(#1)}
\newcommand{\PolER}{\rho_{\mathit{ER}}}
\newcommand{\PfunER}[1]{\PolER(#1)}

\newcommand{\indop}{\mathbbm{1}} 
\newcommand{\indf}[2]{\indop_{#1}\left(#2\right)} 

\newcommand{\Inter}{\cali} 
\newcommand{\Interclique}{\cali^{\textit{clique}}} 
\newcommand{\Interdisconnected}{\cali^{\textit{disc}}} 
\newcommand{\Interfaint}{\cali^{\textit{faint}}} 
\newcommand{\Interunrelenting}{\cali^{\textit{unrel}}} 
\newcommand{\Intermalleable}{\cali^{\textit{malleable}}} 
\newcommand{\Intercircular}{\cali^{\textit{circ}}} 

\newcommand{\Ifun}[2]{\Inter_{#1,#2}} 
\newcommand{\Ifunclique}[2]{\Interclique_{#1,#2}} 
\newcommand{\Ifundisconnected}[2]{\Interdisconnected_{#1,#2}} 
\newcommand{\Ifunfaint}[2]{\Interfaint_{#1,#2}} 
\newcommand{\Ifununrelenting}[2]{\Interunrelenting_{#1,#2}} 
\newcommand{\Ifunmalleable}[2]{\Intermalleable_{#1,#2}} 
\newcommand{\Ifuncircular}[2]{\Intercircular_{#1,#2}} 

\newcommand{\tmax}{T} 

\newcommand{\Upd}{\mu} 
\newcommand{\Ufun}[2]{\Upd(#1,#2)} 

\newcommand{\UpdR}{\mu^{R}} 
\newcommand{\UfunR}[2]{\UpdR(#1,#2)} 

\newcommand{\UpdCB}{\mu^{\textit{CB}}} 
\newcommand{\UfunCB}[2]{\UpdCB(#1,#2)} 


\newcommand{\larrow}[1]{\stackrel{\,\, \small #1\,\,}{\rightarrow}} 
\newcommand{\Larrow}[2]{\stackrel{\,\,#1\,\,}{\leadsto_{#2}}} 
\newcommand{\dinfl}[2]{#1 \larrow{} #2} 
\newcommand{\ldinfl}[3]{#1 \larrow{\begin{tiny}{#2}\end{tiny}} #3} 
\newcommand{\infl}[2]{#1 \Larrow{}{} #2} 
\newcommand{\linfl}[4]{#1 \Larrow{#2}{#3} #4} 

\newcommand{\nat}{\mathbb{N}}
\newcommand{\reals}{\mathbb{R}}
\newcommand{\mstar}{\mathtt{m}}

\newcommand{\defsymbol}{\stackrel{\textup{\texttt{def}}}  {=}} 



\section{Introduction}
\label{sec:introduction}


\todo{Mario on 2020-10-19: Add this to intro somewhere? ``A central concern of our model is that the majority of the (quite extensive) literature on the representation of agents' beliefs --as well as of the way these agents interact to update such beliefs-- assumes that agents are rational and optimally use all information at their disposal. But a growing body of work by psychologists, sociologists, and economists has challenged these assumptions, and several common, relevant cognitive biases  have been catalogued~\cite{XXX,YYY,ZZZ}. In particular, we model the behavior of agents prone to \emph{authority bias}, by which one gives more weight to evidence presented by some agents than by others, and to \emph{confirmation bias}, by which one tends to give more weight to evidence supporting their current beliefs than to  evidence contradicting them, independently from whence the evidence is coming.''}

Social networks facilitate the exchange of opinions by providing users with information from influencers, friends, or other users with similar or sometimes opposing views \cite{Bozdag13}. This may allow a healthy exposure to diverse perspectives, but it may lead also to the problems 
of misinformation and radicalization of opinions.

In social scenarios, a group may shape their beliefs by attributing more value to the opinions of influential figures. This cognitive bias is known as \emph{authority bias} \cite{Ramos:19:Book}. Furthermore, in a group with uniform views, users may become extreme by reinforcing one another's opinions, giving more value to opinions that confirm their own beliefs; another common cognitive bias known as \emph{confirmation bias}~\cite{Aronson10}. As a result, social networks can cause their users to become radical and isolated in their own ideological circle causing dangerous splits in society~\cite{Bozdag13} in a phenomenon known as \emph{polarization}~\cite{Aronson10}.

There is a growing interest in the development of models for the analysis of polarization \cite{li,proskurnikov,sirbu,gargiulo,alexis,Guerra,myp,degroot,naive,zoe,fblogic,facebook,hunter}. To our knowledge, however, the development of \emph{concurrency-based} models for this phenomenon has been far too little considered.  Since polarization involves non-terminating systems with \emph{multiple agents} simultaneously exchanging information (opinions), concurrency models are a natural choice to capture the dynamics of polarization.  

\todo{Mario to Frank on 2020-10-19: Do we want to refer to ourselves as ``the authors'' instead of ``we''?}
In fact, we developed a multi-agent model for polarization in \cite{Alvim:19:FC},  inspired by linear-time models of concurrency where the state of the system evolves in discrete time units (in particular,  \cite{tcc,ntcc}). In each time unit, the agents (users) \emph{update} their beliefs (opinions) according to the underlying influence graph and the beliefs of their neighbors. The \emph{influence graph} is a weighted directed graph describing connectivity and influence of each agent over the others. Two belief updates are considered. The \emph{regular} belief update gives more value to the opinion of agents with higher influence, representing \emph{authority bias}. The \emph{confirmation bias} update  gives more value to the opinion of agents with similar views. Furthermore, the model is equipped with  a \emph{polarization measure} based on  the seminal work in economics by Esteban and Ray~\cite{Esteban:94:Econometrica}.  The polarization is measured at each time unit and it is $0$ if all agents achieve complete consensus about a given proposition. The contributions in \cite{Alvim:19:FC} were of an experimental nature and aimed at exploring how the combination of interaction graphs and cognitive biases in our model can lead to polarization.

In the current paper we prove claims made from experimental observations in \cite{Alvim:19:FC} using techniques from calculus, graph theory, and flow networks. We study meaningful families of influence graphs for which polarization eventually becomes zero (vanishes). In particular we consider cliques, strongly-connected, and balanced-influence graphs, and the belief update can be regular or with confirmation bias. We derive new properties not observed through experiments in \cite{Alvim:19:FC}  and state connections between influence and the notion of flow from network theory. Our results provide insight into the phenomenon of polarization, and are a step toward the design of robust computational models and simulation software for human cognitive and social processes. We therefore believe that FoSSaCS, whose specific topics include models for concurrent systems, would be an excellent venue for this work.
\todo{Mario to Sophia on 2020-10-19: The intro is good! I miss, however, a little
bit more of why we need such a model. We had a good intro on this in the previous paper, maybe we can reuse some of the sentences here?
Something like: ``We believe that our results bring some insights about the phenomenon of polarization, and may help the design of more robust computational models for
human cognitive and social processes.''?
} 
\paragraph{Main Contributions.}  In this paper we establish the following theoretical results:
\begin{enumerate} 
\item  If the influence graph is strongly-connected then the agents' beliefs converge to one value (i.e., they reach consensus), and polarization converges to zero.
\item For weakly-connected and influence-balanced influence graphs (each agent influences others as much as she is influenced)  we give the value of belief convergence. 
\item For clique influence graphs we determine the exact value of belief convergence and the time after which agents reach a given difference of opinion.
\item If polarization does not converge to zero then either there is a group disconnected from the rest, or there is an agent that influences others more than she is influenced. 
\item Polarization does not necessarily converge to zero for weakly-connected graphs. 
\end{enumerate}
We also illustrate a series of case studies and simulations, uncovering interesting new insights and perhaps counter-intuitive characteristics of the phenomena. The code of the implementation of the model and simulations is provided in this paper.
We discuss related work in Section~\ref{sec:related}.
We introduce the model in Section~\ref{sec:model}, and
present case studies and simulations in Section~\ref{sec:simulations}.
The theoretical results are shown in Sections~\ref{sec:specific-cases} and \ref{sec:confirmation-bias}.

\section{Related Work}
\label{sec:related}

As social networks have become ubiquitous, their wide-ranging effects on the world have motivated a great deal of research. 
Here we provide an overview of the range of relevant approaches to the problems we are considering, and put into perspective the novelty of our approach to the problem. 

\noindent\textbf{Polarization} Polarization was originally studied as a psychological phenomenon in \cite{M76},
and it was first rigorously and quantitatively defined by economists Esteban and Ray \cite{Esteban:94:Econometrica}. Their measure of polarization, discussed  in  Section~\ref{sec:model}, is influential, and is the one we adopt in this paper.
Li et al.\cite{li} were the first to model consensus and polarization in social networks. Like most other work, they did not quantify polarization, but focused on  when and under what conditions a population reaches consensus.\todo[fancyline]{Frank: Notice that in this paper that's exactly what we do.} 
Proskurnikov et al. \cite{proskurnikov} investigated the formation of consensus or polarization in social networks, but considered polarization as lack of consensus, rather than a phenomenon 
in its own right. 
Elder's work \cite{alexis} focuses on methods to avoid polarization, without using a quantitative definition of polarization. 
 \cite{Guerra} and \cite{myp} measure polarization but purely as a function of network topology, rather than taking agents' quantitative beliefs and opinions into account.

\noindent\textbf{Formal Models} S{\^\i}rbu et al. \cite{sirbu}
use a model  somewhat similar to ours,
except that it updates probabilistically. 
They investigate the effects of algorithmic bias on polarization by counting the number of opinion clusters, interpreting a single opinion cluster as consensus, rather than directly measuring polarization itself.  
Leskovec et al. \cite{gargiulo} develop simulated social networks and observe group formation over time. Their work is not concerned with a formal measure of polarization.  \cite{Guerra} and \cite{myp} measure polarization but purely as a function of network topology, rather than taking agents' quantitative beliefs and opinions into account. The models in \cite{degroot} and \cite{naive} are closest to ours; however, rather than examining polarization and opinions, this work is concerned with the network topology conditions under which agents with noisy data about an objective fact converge to an accurate consensus, close to the true state of the world. This is related to our goals in the sense that agents who have converged to a consensus have a low level of polarization, but the problem of obtaining accurate information about the world through communication is different than our focus on changing opinions about an issue which does not have an objective, outside value.

%
%
%
%

\noindent\textbf{Logic-based approaches} Liu et al. \cite{liu} use ideas from doxastic and dynamic epistemic logics to
qualitatively model influence and belief change in social networks. 
%
Christoff \cite{zoe} develops several non-quantitative logics for social networks, concerned with problems related to polarization, such as information cascades.
Seligman et al. \cite{fblogic,facebook} introduce a basic ``Facebook logic.'' This logic is non-quantitative, but its interesting point is that an agent's possible worlds are different social networks. This is a promising approach to formal modeling of  epistemic issues in social networks.  
Hunter \cite{hunter} introduces a logic of belief updates over social networks where closer agents in the social network are more trusted and thus more influential. While beliefs in this logic are non-quantitative, it includes a quantitative notion of influence 
between users. 




\section{Background: The Model}
\label{sec:model}
In this section we recall the polarization model introduced in \cite{Alvim:19:FC}. We presuppose basic knowledge of 
calculus and graph theory \cite{Sohrab:14,Diestel:17}.
%
%
%
%
%
We divide the elements of our model into {static} and {dynamic} parts, described next.



%

\subsection{Static elements of the model}
\label{sec:model-static}

\emph{Static elements} of the model represent a snapshot of a social network 
at a given point in time. 
The model includes the following static elements:

\begin{itemize}
 \item A (finite) set $\Agents = \{\agent{0}, \agent{1}, \ldots, \agent{n{-}1} \}$ of $n  \geq 1$ \emph{agents}.
 
 \item A \emph{proposition} $p$ of interest, about which agents can hold beliefs.
 
 \item A \emph{belief configuration}    
    $\Blf{:}\Agents{\rightarrow}[0,1]$ s.t.
    each value $\Blf_{\agent{i}}$ is the instantaneous confidence of agent
    $\agent{i}{\in}\Agents$ in the veracity of proposition $p$.
    The extreme values $0$ and $1$ represent a firm belief in, respectively, the falsehood or truth of the proposition $p$.
 
 \item A \emph{polarization measure} $\Pol{:}[0,1]^{\Agents}{\rightarrow}\R$ 
 mapping belief configurations to real numbers.
 The value $\Pfun{\Blf}$ indicates how polarized belief configuration 
 $\Blf$ is.
\end{itemize}

There are several polarization measures described in the literature.
In this work we concentrate on the influential measure proposed by 
Esteban and Ray~\cite{Esteban:94:Econometrica}.

\begin{definition}[Esteban-Ray polarization measure]
\label{def:poler}
Consider a discrete set $\caly = \{y_0, \allowbreak y_1, \allowbreak \ldots, \allowbreak y_{k-1}\}$ of size $k$, s.t. each $y_i{\in}\mathbb{R}$.
Let $(\pi, y){=}(\pi_0, \allowbreak \pi_1, \allowbreak \ldots, \allowbreak \pi_{k{-}1}, \allowbreak y_0, \allowbreak y_1, \allowbreak \ldots, \allowbreak y_{k{-}1})$ 
be a \emph{distribution} on $\caly$ s.t. $\pi_i$ is the frequency of value $y_i{\in}\caly$ 
in the distribution. 
W.l.o.g. we can assume the values of $\pi_i$ are all non-zero and add up to 1 
(as in a standard probability distribution). 
The \emph{Esteban-Ray (ER) polarization measure} is defined as
$$
    \PfunER{\pi, y} = K \sum_{i=0}^{k-1} \sum_{j=0}^{k-1} \pi_i^{1+\alpha} \pi_j | y_i - y_j |, 
$$
where $K > 0$ is a constant, and typically $\alpha \approx 1.6$.
\end{definition}

The higher the value of $\PfunER{\pi, y}$, the more polarized the 
distribution $(\pi,y)$ is.
The measure captures the intuition that 
polarization is accentuated by both intra-group homogeneity
and inter-group heterogeneity.
Moreover, it assumes that the total polarization 
is the sum of the effects of individual agents on one another.
The measure can be derived from a set of intuitively reasonable
axioms~\cite{Esteban:94:Econometrica}, described in 
Appendix~\ref{sec:polar-axioms}.

Note, however, that $\PolER$ is defined on a discrete distribution, 
whereas in our model a general polarization metric is defined on a 
belief configuration $\Blf{:}\Agents{\rightarrow}[0,1]$. 
To properly apply $\PolER$ to our setup we need to convert the belief 
configuration $\Blf$ into an appropriate distribution $(\pi,y)$ 
as expected by the measure.
We can do this as follows.
Given a number $k{>}0$ of \emph{(discretization) bins}, 
define $\caly{=}\{y_1,y_2,\ldots,y_k\}$ as a discretization of the 
range $[0,1]$ in such a way that each $y_i$ represents the interval
$
	y_i{=}\left[\nicefrac{i}{k},\nicefrac{i+1}{k}\right]
$.
Define, then, the weight $\pi_i$ of value $y_i$, 
corresponding to the fraction of agents having belief in the 
interval $y_{i}$, as
$
	\pi_i{=}\nicefrac{\left(\sum_{a \in \Agents}  \indf{y_i}{\Blf_{\agent{i}}}\right)}{|\Agents|},
$
where $\indf{\calx}{x}$ is the \emph{indicator function} returning 1 if $x \in \calx$, and 0 otherwise. 
Notice that in the particular case in which 
all agents in the belief configuration $\Blf$ hold the same belief value, 
i.e., when there is consensus about the proposition $p$ under consideration,  
$\PolER$ produces 0. 

\subsection{Dynamic elements of the model}
\label{sec:model-dynamic}

\emph{Dynamic elements} of the model capture the  information necessary to formalize the evolution of agents' beliefs in the social network. Although these aspects are not directly used to compute the instantaneous level of polarization, they determine how polarization evolves.
The dynamic elements of our model are the following:

\begin{itemize}
\item An \emph{influence graph} $\Inter{:}\Agents{\times}\Agents {\rightarrow}[0,1]$ 
s.t., for all $\agent{i},\agent{j}{\in}\Agents$, $\Inter(i,j)$, written $\Ifun{\agent{i}}{\agent{j}}$, represents the influence of agent $\agent{i}$ on agent $\agent{j}$. 
A higher value means stronger influence. We shall often refer to $\Inter$ simply as the \emph{influence} $\Inter$. 
\item A \emph{time frame} $\calt{ = }\{0, 1, 2, \ldots \}$ representing the 
discrete passage of time.

\item A \emph{family of belief configurations} $\{\Blft{t}{:}\Agents{\rightarrow}[0,1]\}_{t{\in}{\calt}}$
indexed by time steps s.t. each $\Blft{t}$ is a social network's 
belief configuration w.r.t. proposition $p$ at time step $t{\in}\calt$.

\item  An \emph{update function} $\Upd{:}\Blf{\times}\Inter {\rightarrow}\Blf$ 
mapping a belief configuration and an interaction graph to a new belief configuration.
 $\Blft{t+1}{=}\Ufun{\Blft{t}}{\Inter}$ models the evolution
of agents' beliefs from time step $t$ to $t{+}1$ when their 
interaction is modeled by $\Inter$.
\end{itemize}

The definition of an update function depends
on how agents incorporate new evidence into their reasoning.
In this work the update function considers that the 
impact of agent $\agent{j}$'s belief on agent $\agent{i}$'s is 
proportional to the influence $\Ifun{\agent{j}}{\agent{i}}$ of 
$\agent{j}$ on $\agent{i}$, and to the difference 
$\Bfun{\agent{j}}{t}{ - }\Bfun{\agent{i}}{t}$ in their beliefs.
The function then averages all influences on an agent to compute
the corresponding belief update. 
This intuition is formalized next.

\begin{definition}[Update function]
    \label{def:rational-update-function}
    Let $\Blft{t}$ be a belief configuration at time step $t{\in}\calt$,
    and $\Inter$ be an influence graph.
    The \emph{update function} $\UpdR{:}\Blf{\times}\Inter{\rightarrow}\Blf$ 
    is defined s.t. the updated belief configuration $\Blft{t+1}{=}\UfunR{\Blft{t}}{\Inter}$ at time step $t{+}1$, for all agents $\agent{i}$, 
    is given by
    \begin{align*}
        \Bfun{\agent{i}}{t+1} = \frac{1}{| \Agents |} \sum_{\agent{j} \in \Agents} \Bapp{\agent{i}}{\agent{j}}{t+1}~, 
    \quad
    \text{where} 
    \quad 
        \Bapp{\agent{i}}{\agent{j}}{t+1} = \Bfun{\agent{i}}{t} + \Ifun{\agent{j}}{\agent{i}} \left( \Bfun{\agent{j}}{t} - \Bfun{\agent{i}}{t} \right)~
    \end{align*}
    represents the influence of agent $\agent{j}\in\Agents$
    on agent $\agent{i}$'s belief at time $t{+}1$ after 
    both agents interact at time $t$.
    The value $\Bfun{\agent{i}}{t+1}$ is agent $\agent{i}$'s 
    belief at time $t{+}1$, resulting from the weighted average
    of interactions with all agents at time $t$, in parallel.
\end{definition}

\todo{Mario on 2020-10-19: Discussion about bias added in this paragraph.}
Notice that the above update function models 
agents who incorporate equitably all evidence available, 
be it in favor or against a proposition, when updating their beliefs.
Yet, influence graphs allow us to capture different intensities of \emph{authority bias}~\cite{Ramos:19:Book}, by which an agent 
gives more weight to evidence presented by some agents than by others.\footnote{In Section~\ref{sec:confirmation-bias} we consider an extension of the model capturing agents prone to \emph{confirmation bias}, by which one tends 
to give more weight to evidence supporting their current beliefs than to 
evidence contradicting them, independently from whence the evidence is coming.}
As a particular case, a proper setting of influence values can capture 
agents sensitive only to the contents of information, independent 
from its source.

\section{Simulations and Motivating Examples}
\label{sec:simulations}

In this section we present several simulations
of the evolution of beliefs and polarization in our model,
employing various combinations of meaningful influence graphs 
and initial belief configurations.
These examples provide insights that are 
captured as formal properties in Sec.\ref{sec:specific-cases}.

For the computation of the Esteban-Ray measure (Def.~\ref{def:poler}), 
we discretize the interval $[0,1]$ of possible belief values 
into 5 bins, each representing a possible general position 
w.r.t. the veracity of the proposition $p$ of interest:
\textit{strong disagreement}, $[0,0.20)$;
\textit{mild disagreement}, $[0.20,0.40)$;
\textit{indifference}, $[0.40,0.60)$;
\textit{mild agreement}, $[0.60,0.80)$; and
\textit{strong agreement}, $[0.80,1]$.
We set parameters $\alpha{=}1.6$, 
as suggested by Esteban and Ray~\cite{Esteban:94:Econometrica},
and $K{=}1000$.
In all definitions we let $\Agents{=}\{0, 1,\ldots, n{-}1 \}$, 
and $\agent{i},\agent{j}{\in}\Agents$ be generic agents.
Each simulation uses a set of $n{=}100$ agents
(unless otherwise noted), and limits execution to 
a threshold of $\tmax$
time steps, varying according to the experiment.

\begin{figure}[tb]
    \centering
    \includegraphics[width=0.85\linewidth]{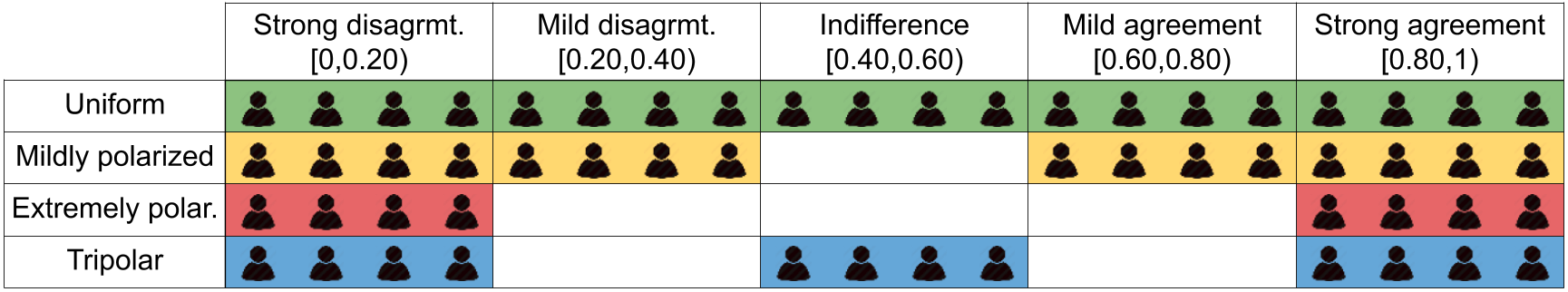}
    \caption{Depiction of different 
    initial belief configurations used in simulations.
    }
    \label{fig:initial-belief-configurations}
\end{figure}

We consider the following initial belief configurations,
depicted in Figure~\ref{fig:initial-belief-configurations}.

\begin{itemize}
    \item A \emph{uniform} belief configuration representing a set of agents whose beliefs are as varied as possible, all equally spaced in the interval $[0, 1]$: 
    $\Bfun{i}{0}{=}\nicefrac{i}{(n{-}1)}.$
    
    \item A \emph{mildly polarized} belief configuration representing agents evenly split into two groups with moderately dissimilar inter-group beliefs compared to their intra-group beliefs: 
    $\Bfun{i}{0}{=}\nicefrac{0.8 i}{n}$, 
    if $i{<}\ceil{\nicefrac{n}{2}}$; or 
    $\Bfun{i}{0}{=}\nicefrac{0.8 i}{n}{+}0.20$, otherwise.
    \item An \emph{extremely polarized} belief configuration
    representing a situation in which half of the 
    agents strongly believe the proposition, whereas the 
    other half strongly disbelieve it:
    $\Bfun{i}{0}{=}\nicefrac{0.4 i}{n}$, 
    if $i{<}\ceil{\nicefrac{n}{2}}$; or 
    $\Bfun{i}{0}{=}\nicefrac{0.4 i}{n}{+}0.60$, otherwise.
    
    \item A \emph{tripolar} configuration representing 
    agents divided into three groups of similar size sharing
    similar beliefs:
    $\Bfun{i}{0}{=}\nicefrac{0.60 i}{n}$, if $0{\leq} i{<} \lfloor\nicefrac{n}{3}\rfloor$; 
    $\Bfun{i}{0}{=}\nicefrac{0.60 i}{n}+0.20$, if $\lfloor\nicefrac{n}{3}\rfloor{\leq} i{<} \ceil{\nicefrac{2n}{3}}$; or $\Bfun{i}{0}{=}\nicefrac{0.60 i}{n}+0.40$, otherwise.
\end{itemize}

\begin{figure}[tb]
    \centering
    \begin{subfigure}[t]{0.15\textwidth}
      \centering
      \includegraphics[width=\textwidth]{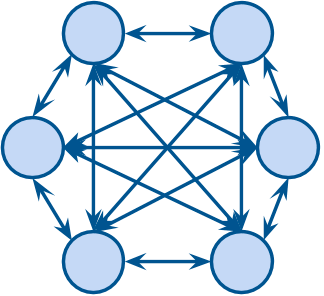}
      \caption{Clique}
      \label{fig:interaction-graphs-clique}
    \end{subfigure}
    \hfill
    \begin{subfigure}[t]{0.15\textwidth}
      \centering
      \includegraphics[width=\textwidth]{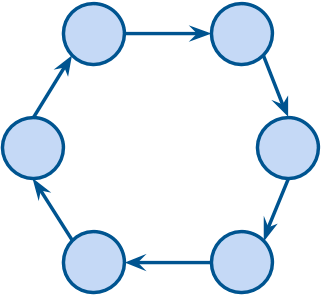}
      \caption{Circular}
      \label{fig:interaction-graphs-circular}
    \end{subfigure}
    \hfill
    \begin{subfigure}[t]{0.30\textwidth}
      \centering
      \includegraphics[width=0.9\textwidth]{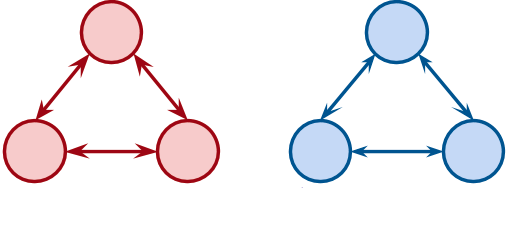}
      \caption{Disconnected groups}
      \label{fig:interaction-graphs-disconnected}
    \end{subfigure}
    \hfill
    \begin{subfigure}[t]{0.33\textwidth}
      \centering
      \includegraphics[width=0.8\textwidth]{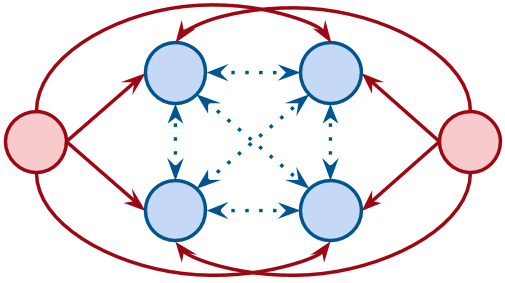}
      \caption{Unrelenting influencers}
      \label{fig:interaction-graphs-unrelenting}
    \end{subfigure}
    \caption{The general shape of influence graphs used in simulations.}
    \label{fig:interaction-graphs}
\end{figure}

As for influence graphs, we consider the following ones,
depicted in Figure~\ref{fig:interaction-graphs}.

\begin{itemize}
    
    \item A \emph{$C$-clique} influence graph $\Interclique$
    (formalized in Def.~\ref{def:c-clique} ahead), 
    in which each agent influences every other with constant 
    value $C=0.5$: $\Ifunclique{\agent{i}}{\agent{j}}{=}0.5$, 
    for every $\agent{i},\agent{j}$.
    \todo{Mario on 2020-10-19: Added this.}
    This represents the particular case of a social network in
    which all agents interact among themselves, and are all immune
    to authority bias.
    
    \item A \emph{circular} influence graph $\Intercircular$ 
    representing a social network in which agents can be organized
    in a circle in such a way each agent is only influenced by its 
    predecessor and only influences its successor.
    This is a simple instance of a balanced graph (in which
    each agent's influence on others is as high as the influence received, 
    as in Def.~\ref{def:circulation} ahead), which is a pattern commonly 
    encountered in some sub-networks.
    More precisely:
    $\Ifuncircular{\agent{i}}{\agent{j}}{=}0.5$, if $(i{+}1)\,\text{mod}\,n{=}j$; and $\Ifuncircular{\agent{i}}{\agent{j}}{=}0$, otherwise.
    
    \item A \emph{disconnected} influence graph $\Interdisconnected$
    representing a social network sharply divided into two groups in such a way that agents within the same group 
    can considerably influence each other, but not at all
    agents in the other group:
    $\Ifundisconnected{\agent{i}}{\agent{j}}{=}0.5$, if $\agent{i},\agent{j}$ are both
    ${<} \ceil{\nicefrac{n}{2}}$ or both ${\geq} \ceil{\nicefrac{n}{2}}$, 
    and $\Ifundisconnected{\agent{i}}{\agent{j}}{=}0$ otherwise.

    
    \item An \emph{unrelenting influencers} influence graph $\Interunrelenting$ 
    representing a scenario in which two agents 
    (say, $\agent{0}$ and $\agent{n{-}1}$) 
    exert significantly stronger influence on every other agent than these other agents have among themselves.
    More precisely:
    $\Ifununrelenting{\agent{i}}{\agent{j}}{=}0.6$ if $i{=}0$ and $j{\neq}n{-}1$ or $i{=}n{-}1$
    and $j{\neq}0$;
    $\Ifununrelenting{\agent{i}}{\agent{j}}{=}0$ if 
    $j{=}0$ or $j{=}n{-}1$;
    and finally
    $\Ifununrelenting{\agent{i}}{\agent{j}}{=}0.1$ if $0{\neq} i{\neq}n{-}{1}$ and $0{\neq}j{\neq}n{-}{1}$.
    This could represent, e.g., a social network 
    in which two totalitarian media companies dominate the news
    market, both with similarly high levels of influence on 
    all agents.
    The networks have clear agendas to push forward, and are not influenced in a meaningful way by other agents.
      
\end{itemize}

\begin{figure}[htp]
\centering
  \includegraphics[width=\linewidth]{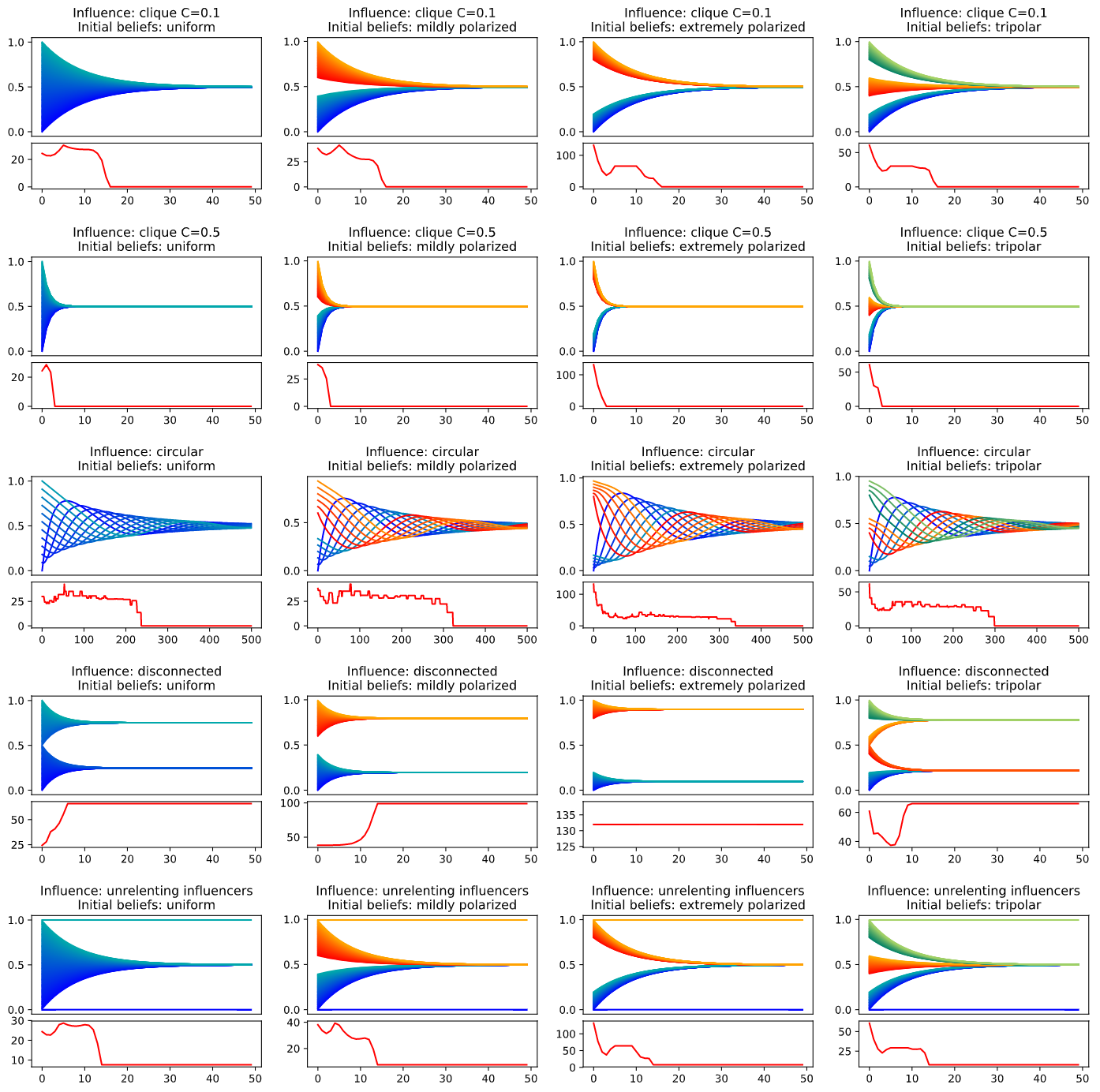}
  \caption{Simulations of the evolution of beliefs (upper part of each graph) and of polarization 
  (lower part of each graph) under different initial belief configurations 
  and influence graphs. 
  In each graph the horizontal axis represents the discrete passing of time.
  Each row contains all graphs with the same influence graph, and each column 
  all graphs with the same initial belief configuration.
  All simulations considered $n{=}100$ agents, with the exception of those regarding the circular influence graph, which used $n{=}12$ agents.
  }
  \label{fig:comparing-num-bins}
\end{figure}

We simulated the evolution of agents' beliefs
and the corresponding polarization of the network 
for all combinations of initial belief configurations 
and influence graphs presented above.
The results, depicted in Figure~\ref{fig:comparing-num-bins},
will be used throughout this paper to illustrate some of our 
formal results. 
Both the Python implementation of the model and the Jupyter Notebook 
containing the simulations are available on Github~\cite{website:github-repo}.
\todo{Santiago: Frank, I added a reference to polar.bib with the repository.}


\section{Consensus under Strongly-Connected Influence}
\label{sec:specific-cases}


 Polarization eventually disappears if the agents asymptotically agree upon a common belief value for the proposition of interest; i.e., if they reach \emph{consensus}.  In our model this means that the polarization measure (Def.\ref{def:poler}) converges to zero  if  the limiting value of  agent beliefs is the same as time approaches infinity. 
 
 In this section, the main and most technical  of this paper, we consider meaningful families of influence graphs that guarantee consensus for all agents.
 We also identify fundamental properties of agents under this influence,
 as well as the time to achieve a given opinion difference and the actual value of convergence. We shall also relate influence with the notion of flow in flow networks. 

\subsection{Constant Cliques}



%
%
%
%

We now introduce the notion of clique influence graph. It is meant to capture a scenario 
where all agents influence one another with the same value.

\begin{definition}[Clique Influence]
\label{def:c-clique}
 Let $C$ be a real value in $(0,1].$ We say that the influence graph $\Inter$ is a  \emph{$C$-clique} if for all $\agent{i}, \agent{j} \in \Agents$ if $i \neq j$ then $\Ifun{\agent{i}}{\agent{j}} = C.$ 
\end{definition}
The following equations are immediate consequences of restricting  Def.\ref{def:rational-update-function}  to  $C$-cliques.

\begin{eqnarray}
\Bfun{i}{t+1} 
\quad=\quad 
\Bfun{i}{t} + \frac{C}{\mid\Agents\mid} \sum_{\agent{j} \in \Agents} \left( \Bfun{j}{t} - \Bfun{i}{t} \right) 
\quad=\quad  
\Bfun{i}{t} - C \Bfun{i}{t} + \frac{C}{\mid\Agents\mid} \sum_{\agent{j} \in \Agents} \Bfun{j}{t} \label{eq2} \label{eq3}
\end{eqnarray}


The next definition identifies agents with \emph{extreme beliefs} at a given time. 

\begin{definition}[Extreme Agents]\label{extreme:def} We say that an agent 
$\agent{i}\in\Agents$ is  \emph{maximal (minimal)} at time $t$ if $\Bfun{i}{t} \geq \Bfun{j}{t}$ ($\Bfun{i}{t} \leq \Bfun{j}{t}$) for all $\agent{j} \in \Agents$. An \emph{extreme agent} at time $t$ is an agent that is maximal or minimal at time $t$. 
\end{definition}
%


 An invariant property of extreme agents under clique influence is that they cannot become more extreme.

\begin{proposition}
\label{prop:max-min-monotone}
If $\Inter$ is a  {$C$-clique} and $\agent{i} \in \Agents$ is \emph{maximal (minimal)} at time $t$ then
\begin{enumerate}
 \item $\Bfun{i}{t+1} \leq \Bfun{i}{t}$ ($\Bfun{i}{t+1} \geq \Bfun{i}{t}$).
 \item  $ \Bfun{i}{t+1} <\Bfun{i}{t}$    ($\Bfun{i}{t+1} > \Bfun{i}{t}$) if there is $\agent{j}\in\Agents$ such that $\Bfun{j}{t} \neq \Bfun{i}{t}.$
 \end{enumerate}
\end{proposition}
\begin{proof}
Let $\agent{i}\in \Agents$ be maximal agent at time $t$. (1) We have  $\sum_{\agent{j} \in \Agents} \left( \Bfun{j}{t} - \Bfun{i}{t} \right) \leq 0.$
Hence from Eq.\ref{eq2}, $ \Bfun{i}{t+1} = \Bfun{i}{t} + \nicefrac{C}{\mid\Agents\mid} \sum_{\agent{j} \in \Agents} \left( \Bfun{j}{t} - \Bfun{i}{t} \right)\leq \Bfun{i}{t}.$  (2) Assume $\agent{j}\in\Agents$ with $\Bfun{j}{t} \not= \Bfun{i}{t}.$ Then $\sum_{\agent{j} \in \Agents} \left( \Bfun{j}{t} - \Bfun{i}{t} \right) < 0$ and thus, from Eq.\ref{eq3},  $ \Bfun{i}{t+1} = \Bfun{i}{t} + \nicefrac{C}{\mid\Agents\mid} \sum_{\agent{j} \in \Agents} \left( \Bfun{j}{t} - \Bfun{i}{t} \right) < \Bfun{i}{t}. $ 
The cases when $i$ is minimal are similar. \qed
\end{proof}

A distinguishing property for $C$-cliques is that extreme agents continue to  be extreme. This property is a consequence of the following proposition.

\begin{proposition}[Order Preservation]\label{prop:clique-belief-order} Assume that the influence graph $\Inter$ is a  {$C$-clique}.  For any $\agent{i}, \agent{j} \in\Agents$, if $\Bfun{i}{t} \geq \Bfun{j}{t}$ then $\Bfun{i}{t+1} \geq \Bfun{j}{t+1}.$
\end{proposition}

\begin{proof} Assume $\Bfun{i}{t}{\geq}\Bfun{j}{t}$. From Eq.\ref{eq3} we have 
$\Bfun{i}{t+1}{=}\Bfun{i}{t}{-}C \Bfun{i}{t} {+} \nicefrac{C}{\mid\Agents\mid} \sum_{\agent{k} \in \Agents} \Bfun{k}{t}$ and $\Bfun{j}{t+1} = \Bfun{j}{t} - C \Bfun{j}{t} + \nicefrac{C}{\mid\Agents\mid} \sum_{\agent{k} \in \Agents} \Bfun{k}{t}.$ It follows that $\Bfun{i}{t+1} - \Bfun{j}{t+1} = (1-C) (\Bfun{i}{t} - \Bfun{j}{t}).$ Since $C \in [0,1]$ and $\Bfun{i}{t} \geq \Bfun{j}{t}$, $\Bfun{i}{t+1} - \Bfun{j}{t+1} \geq 0$, thus  $\Bfun{i}{t+1} \geq \Bfun{j}{t+1}.$ \qed 
\end{proof}

%

The next property follows from  Prop. \ref{prop:clique-belief-order} and Prop. \ref{prop:max-min-monotone}, and states that the difference of opinion among extreme agents in $C$-cliques cannot grow bigger.
\begin{proposition}[Decreasing Opinion Difference]\label{prop:decreasing-difference}
Suppose that $\Inter$ is a  {$C$-clique}. If $\agent{i},\agent{j} \in \Agents$ are maximal and minimal agents at time $t$ then $\Bfun{i}{t+1} - \Bfun{j}{t+1} \leq \Bfun{i}{t} - \Bfun{j}{t}$.
\end{proposition}

The next result states that the sum of beliefs in a {$C$-clique} 
remains constant. 

\begin{proposition}[Belief Conservation]\label{prop:const-mean}
Assume that the influence graph $\Inter$ is a  {$C$-clique}.  For every $t>0$ we have $\sum_{\agent{i} \in \Agents} \Bfun{i}{t} = \sum_{\agent{i} \in \Agents} \Bfun{i}{0}.$
\end{proposition}

\begin{proof} It suffices to show that  $\sum_{\agent{i} \in \Agents} \Bfun{i}{t} = \sum_{\agent{i} \in \Agents} \Bfun{i}{t-1}.$
From Eq. \ref{eq3} $\sum_{\agent{i} \in \Agents} \Bfun{i}{t} = \sum_{\agent{i} \in \Agents} \left( \Bfun{i}{t-1} - C \Bfun{i}{t-1} + \nicefrac{C}{\mid\Agents\mid} \sum_{\agent{j} \in \Agents} \Bfun{j}{t-1}\right)$. Expanding the right hand side we then obtain $\sum_{\agent{i} \in \Agents} \Bfun{i}{t} =\sum_{\agent{i} \in \Agents} \Bfun{i}{t-1} - C\sum_{\agent{i} \in \Agents} \Bfun{i}{t-1} + C\sum_{\agent{j} \in \Agents} \Bfun{j}{t-1}=\sum_{\agent{i} \in \Agents} \Bfun{i}{t-1}.$ \qed
\end{proof}

In every $C$-clique the difference of opinion between any two agents decreases by a factor determined by time and the influence $C$. This is formally stated next. 

\begin{proposition}\label{prop:inv-diff}
If $\Inter$ is  {$C$-clique} and  $n\geq 1$ then
$\Bfun{i}{n} - \Bfun{j}{n} = (1-C)^n (\Bfun{i}{0} - \Bfun{j}{0}).$ 
\end{proposition}
\begin{proof} From the proof of Prop.\ref{prop:clique-belief-order}, $\Bfun{i}{n} - \Bfun{j}{n} = (1-C)(\Bfun{i}{n-1} - \Bfun{j}{n-1})$.
The rest of the proof proceeds by an easy induction on $n$.
\qed
\end{proof}
The following lemma states that extreme agents converge to the same value. It also tells us, given a real value $\epsilon$, a time $T_\epsilon$ after which the difference of opinion by extreme agents becomes smaller than $\epsilon$. 

\begin{lemma}[Convergence Time]\label{prop:max-min-order-convergence}
Assume that the influence graph $\Inter$ is a  {$C$-clique}. 
Suppose that  $\agent{i},\agent{j} \in \Agents $ are, respectively, maximal and minimal at time $0.$ For  $\epsilon \in \reals^+,$ define $T_\epsilon=1$ if $C=1$ else $T_\epsilon=\log_{1-C}\left(\nicefrac{\epsilon} {(\Bfun{i}{0} - \Bfun{j}{0})}\right).$  
\begin{multicols}{3}
\begin{enumerate} 
    \item $t{>}T_\epsilon \Rightarrow \Bfun{i}{t}{-}\Bfun{j}{t}{<}\epsilon.$
    \item $ \underset{t\to\infty}{\lim} (\Bfun{i}{t} - \Bfun{j}{t}) = 0.$
    \item $\Bfun{j}{t}{\leq}\nicefrac{1}{\mid\Agents\mid}\sum_{\agent{k} \in \Agents} \Bfun{k}{0}{\leq}\Bfun{i}{t}.$
\end{enumerate}
\end{multicols}
\end{lemma}

\begin{proof}
Suppose that  $\agent{i}, \agent{j} \in\Agents$ are the maximal and minimal agents at time $0$.
\begin{enumerate}
    
%
%

   \item   Let  $\epsilon > 0$ and $t$ be arbitrary real and natural numbers. 
   If  $C=1$ from Eq.\ref{eq3} if $t>1$, $\Bfun{i}{t} = \Bfun{j}{t} = \nicefrac{1}{\mid\Agents\mid} \sum_{\agent{k} \in \Agents} \Bfun{k}{t-1}.$ Thus  if $t > T_\epsilon =1 $ then $\Bfun{i}{t} - \Bfun{j}{t}=0 < \epsilon.$ 
   
   Assume $C<1$. From Prop.\ref{prop:inv-diff} 
   for $n \geq 1$ we have
   $\Bfun{i}{n} - \Bfun{j}{n} = (1-C)^n (\Bfun{i}{0} - \Bfun{j}{0})$. Notice that $(1-C)^n < 1$ for every $n\geq 1$. Thus let $m$ be the smallest natural number such that $(1-C)^m (\Bfun{i}{0} - \Bfun{j}{0}) < \epsilon .$ Clearly $m$ is the smallest natural number such that $m >   T_\epsilon = \log_{1-C}\left(\epsilon / (\Bfun{i}{0} - \Bfun{j}{0})\right)$. Thus  $\Bfun{i}{m} - \Bfun{j}{m} = (1-C)^m (\Bfun{i}{0} - \Bfun{j}{0}) < \epsilon$ and if $t> T_\epsilon$ then $t \geq m$. It follows from Prop.\ref{prop:decreasing-difference} that if $t> T_\epsilon$ we have $\Bfun{i}{t} - \Bfun{j}{t} \leq \Bfun{i}{m} - \Bfun{j}{m}   < \epsilon$ as wanted.
   
\item It is an immediate consequence of  (1).
        \item From Prop.\ref{prop:clique-belief-order}, $\agent{i}$ is  {maximal} and $\agent{j}$  {minimal} at every time $t$. Hence we obtain  $\Bfun{j}{t} \leq \nicefrac{1}{\mid\Agents\mid} \sum_{\agent{k} \in \Agents} \Bfun{k}{t} \leq \Bfun{i}{t}$ for every $t.$ From Prop.\ref{prop:const-mean},  $\sum_{\agent{k} \in \Agents} \Bfun{k}{t} = \sum_{\agent{k} \in \Agents} \Bfun{k}{0}$ for any $t.$ Therefore, $\Bfun{j}{t} \leq \nicefrac{1}{\mid\Agents\mid} \sum_{\agent{j} \in \Agents} \Bfun{k}{0} \leq \Bfun{i}{t}$ for any $t.$ 
        \qed
\end{enumerate}
\end{proof}

Using the previous lemma we obtain
the opinion value to which all agents converge in a $C$-clique: 
the average of all initial opinions.  

\begin{theorem}[Consensus Value]\label{thm:classcliq-convergence}
Assume that the influence graph $\Inter$ is a  {$C$-clique}. For every $i\in \Agents$, 
 $ \lim_{t\to\infty}\Bfun{i}{t} = \nicefrac{1}{\mid\Agents\mid} \sum_{\agent{k} \in \Agents} \Bfun{k}{0}. $ 
\end{theorem}
\begin{proof} An immediate consequence of Lem.\ref{prop:max-min-order-convergence} and the squeeze theorem \cite{Sohrab:14}. \qed
\end{proof}

Notice that Th.\ref{thm:classcliq-convergence} says that for $C$-cliques the opinion convergence  value is independent of the influence $C$. Nevertheless,  Lem.\ref{prop:max-min-order-convergence} tells us that the smaller the value of $C$, the longer the time to  reach a given opinion difference $\epsilon$. We conclude with an example illustrating the above mentioned results.

\begin{example}[Beliefs convergence in $C$-cliques] 
The first row of results in 
Fig.~\ref{fig:comparing-num-bins}
shows the evolution of belief and polarization, 
under a variety of initial belief configurations, 
for a $C$-clique with $C{=}0.1$.
The second row does the same for a $C$-clique with $C{=}0.5$.
In all these simulations the belief of every agent converges to the average, but convergence is faster for the $C$-clique with higher influence.
More precisely, since in all initial configurations considered 
the minimal and maximum initial beliefs are 0 and 1, 
respectively, in all cases the $C$-clique 
with $C{=}0.1$ achieves a convergence up to $\epsilon{=}10^{-2}$ in less than 
44 steps, whereas  the clique with $C{=}0.5$ does so in less than 7 steps.
\end{example}


\subsection{Strongly Connected Influence}




We now introduce the family of influence graphs, which includes cliques, that describes scenarios where each agent has an influence over all others. Such influence is not necessarily \emph{direct} in the sense defined next, or the same for all agents, as in the more specific cases of cliques. We call this family \emph{strongly connected} influence graphs. 

\begin{definition}[Influence Paths] \label{def:influence-path} Let $C \in (0,1].$
We say that $\agent{i}$ has a \emph{direct influence} $C$ over $\agent{j}$, written  $\ldinfl{\agent{i}}{C}{\agent{j}}$, if $\Ifun{\agent{i}}{\agent{j}}=C.$ 
 
 An \emph{influence path} is a {finite sequence} of \emph{distinct} agents from $\Agents$ where each agent in the sequence has a direct influence over the next one. Let  $p$ be an influence path $i_0i_1\ldots i_n.$ The \emph{size} of $p$ is $|p|=n$. 
We also use $\ldinfl{\agent{i_0}}{C_1}{\agent{i_1}}\ldinfl{}{C_2}{}\ldots \ldinfl{}{C_n} {\agent{i_n}}$ to denote  $p$ with the direct influences along this path. We write  $\linfl{\agent{i_0}}{C}{p}{\agent{i_n}}$ to indicate that the \emph{product influence} of $i_0$ over $i_n$ along  $p$ is $C=C_1 \times \ldots \times C_n$. 

 We often omit influence or path indices from the above arrow notations when they are unimportant or clear from the context.  We say that   $\agent{i}$ has an \emph{influence} over $\agent{j}$ if $\infl{\agent{i}}{\agent{j}}$. 
\end{definition}

The following definition is reminiscent of the graph theoretical notion of strongly-connected graphs. 

\begin{definition}[Strongly Connected Influence]  We say that an influence graph $\Inter$ is \emph{strongly connected} if for all $\agent{i}$, $\agent{j} \in \Agents$ if $i\neq j$ then $\infl{\agent{i}}{\agent{j}}$.
\end{definition}



We shall use the notion of maximum and minimum belief values at a given time $t$. They are the belief values of extreme agents at $t$ (see Def.\ref{extreme:def}).

\begin{definition}[Extreme Beliefs]\label{def:extreme:beliefs} Define   $\mx{t}=\underset{\agent{i} \in \Agents}{\max} \ \Bfun{\agent{i}}{t}$ and $\mn{t}= \underset{\agent{i} \in \Agents}{\min} \ \Bfun{\agent{i}}{t}.$
 \end{definition}
 
 Recall that in the more specific case of cliques, the extreme agents remain the same across time units (Prop.\ref{prop:clique-belief-order}). Nevertheless \emph{this is not necessarily} the case for strongly connected influence graphs. In fact,  belief order preservation, stated for cliques in  Prop.\ref{prop:clique-belief-order}, does not hold in general for strongly connected influence. 
 
\begin{example}[Non-preservation of belief order] Consider 
the third row of simulation results in Fig.~\ref{fig:comparing-num-bins}, 
which depicts the evolution of belief and polarization
for a circular graph under a variety of initial belief configurations.
This influence is strongly connected, but some agents 
influence others only indirectly.
In these simulations, under all initial belief configurations
there is no order preservation in beliefs.
In fact, different agents alternate continuously as maximal 
and minimal belief holders.
\end{example}

Nevertheless, we will show that the beliefs of all agents under strongly-connected influence converge to the same value since the difference between $\mn{t}$ and $\mx{t}$ goes to 0 as $t$ approaches infinity.  
We use the following equation, derived from Def.\ref{def:rational-update-function}. 

\begin{equation}\label{eq:symplification}
    \Bfun{\agent{i}}{t{+}1} =  \Bfun{\agent{i}}{t} + \frac{1}{|\Agents|}\sum_{\agent{j} \in \Agents}\Ifun{\agent{j}}{\agent{i}}(\Bfun{\agent{j}}{t} -\Bfun{\agent{i}}{t})
\end{equation}



The following lemma states a distinctive property of strongly connected influence: {The belief value of any agent at any time is bounded by those from extreme agents in the previous time unit}. 

\begin{lemma}[Belief Bounds] \label{lemma:maxdiffmin} Assume that the influence graph $\Inter$ is strongly connected. Then  for every $i\in \Agents$,   $\mn{t} \leq \Bfun{\agent{i}}{t{+}1} \leq \mx{t}.$
\end{lemma}
\begin{proof}
%
We want to prove $\Bfun{\agent{i}}{t{+}1} \leq \mx{t}$. Since $\Bfun{\agent{j}}{t} \leq \mx{t}$, we can use Eq.\ref{eq:symplification} 
to derive the inequality $\Bfun{\agent{i}}{t{+}1} 
    \leq E_1\defsymbol\Bfun{\agent{i}}{t} + \nicefrac{1}{|\Agents|}\sum_{\agent{j} \in \Agents}  \Ifun{\agent{j}}{\agent{i}}(\mx{t} -\Bfun{\agent{i}}{t}).$ Furthermore, $E_1 \leq E_2 \defsymbol\Bfun{\agent{i}}{t} + \nicefrac{1}{|\Agents|}\sum_{\agent{j} \in \Agents} (\mx{t} -\Bfun{\agent{i}}{t})$ because $\Ifun{\agent{j}}{\agent{i}} \leq 1$ and $\mx{t} - \Bfun{\agent{i}}{t} \geq 0.$ We thus obtain $\Bfun{\agent{i}}{t{+}1} \leq E_2 = \Bfun{\agent{i}}{t} + \nicefrac{|\Agents|}{|\Agents|}(\mx{t} -\Bfun{\agent{i}}{t}) = \Bfun{\agent{i}}{t} + \mx{t} -\Bfun{\agent{i}}{t} = \mx{t}$ as wanted. The proof of $\mn{t} \leq \Bfun{\agent{i}}{t{+}1}$ is similar. \qed
\end{proof}

   As an immediate consequence of the above lemma, the next corollary tells us that $\mn{\cdot}$ and $\mx{\cdot}$ are monotonically increasing and decreasing functions. 
    
\begin{corollary}[Monotonicity of Extreme Beliefs]\label{cor:mbefore-mafter} If the influence graph $\Inter$ is strongly connected then 
  $\mx{t+1} \leq \mx{t}$ and $min^{t+1} \geq \mn{t}$ for all $t\in\nat$.
  \end{corollary}
  
 \begin{example}[Non-monotonicity of non-extreme beliefs] 
 Note that monotonicity does not necessarily hold for non-extreme beliefs. 
 In all simulations on circular influence graphs, 
 shown in the third row of Fig.~\ref{fig:comparing-num-bins}, 
 it is possible to see that several agents have beliefs evolving 
 in a non-monotonic way.
 \end{example}

%

The above monotonicity property and the fact that both $\mx{\cdot}$ and $\mn{\cdot}$ are bounded between $0$ and $1$ lead us, via the monotone convergence theorem \cite{Sohrab:14}, 
to the existence of \emph{limits for the beliefs of extreme agents}.

\begin{corollary}[Limits of Extreme Beliefs]\label{cor:max-limits-exist} If the influence graph $\Inter$ is strongly connected then 
    $\lim_{t\to\infty} \mx{t} = U$ and $\lim_{t\to\infty} \mn{t} = L$ for some $U$, $L \in [0,1]$.
\end{corollary}

%

Another distinctive property of agents under strongly component influence is that the belief of any agent at time $t$ will influence every other agent by the time $t+|\Agents|-1$. This will be derived from the path influence lemma below (Lem.\ref{lemma:path:influence}). First we need the following rather technical proposition to prove the lemma. 
\begin{proposition}\label{prop:upper-bound-inequality} Assume that
$\Inter$ is strongly connected. Let $i,k \in \Agents$, $n,t \in \nat$ with $n\geq 1$.
   \begin{enumerate} 
\item  If $\Bfun{\agent{i}}{t} \leq v $ then 
    $\Bfun{\agent{i}}{t{+}1} \leq v + \nicefrac{1}{|\Agents|}\sum_{\agent{j} \in \Agents}\Ifun{\agent{j}}{\agent{i}}\left(\Bfun{\agent{j}}{t}-v\right).$ 
 \item $\Bfun{\agent{i}}{t+n} \leq \mx{t} + \nicefrac{1}{|\Agents|}\Ifun{\agent{k}}{\agent{i}}(\Bfun{\agent{k}}{t+n-1} - \mx{t}).$ 
     \end{enumerate}
\end{proposition}
\begin{proof} Let $i,k \in \Agents$, $n,t \in \nat$ with $n\geq 1$.
\begin{enumerate}
\item   From Def.\ref{def:rational-update-function} we obtain $\Bfun{\agent{i}}{t{+}1}  =  \nicefrac{1}{|\Agents|}\sum_{\agent{j} \in \Agents} \left(\Bfun{\agent{i}}{t}(1-\Ifun{\agent{j}}{\agent{i}}) + \Ifun{\agent{j}}{\agent{i}}\Bfun{\agent{j}}{t}\right)$. If $\Bfun{\agent{i}}{t} \leq v$ then $\Bfun{\agent{i}}{t{+}1} \leq  \nicefrac{1}{|\Agents|}\sum_{\agent{j} \in \Agents} \left(v(1-\Ifun{\agent{j}}{\agent{i}}) + \Ifun{\agent{j}}{\agent{i}}\Bfun{\agent{j}}{t}\right) =v + \nicefrac{1}{|\Agents|}\sum_{\agent{j} \in \Agents} \Ifun{\agent{j}}{\agent{i}}\left(\Bfun{\agent{j}}{t} - v\right)$.

\item By Def.\ref{def:rational-update-function} $\Bfun{\agent{i}}{t{+}n}{{=}}\nicefrac{1}{|\Agents|}\sum_{\agent{j} {\in} \Agents}\left(\Bfun{\agent{i}}{t{+}n{-}1} {+}   \Ifun{\agent{j}}{\agent{i}}(\Bfun{\agent{j}}{t{+}n{-}1} {-}\Bfun{\agent{i}}{t{+}n{-}1})\right).$ From Cor.\ref{cor:mbefore-mafter} we have $\Bfun{\agent{i}}{t{+}n} {\leq} \mx{t{+}n} {\leq} \mx{t{+}n{-}1}$. Thus by Prop.\ref{prop:upper-bound-inequality}(1) we have:  
$
        \Bfun{\agent{i}}{t{+}n} 
        \leq \allowbreak \nicefrac{1}{|\Agents|} \cdot \allowbreak \sum_{\agent{j} {\in} \Agents}\left(\mx{t{+}n{-}1} {+}   \Ifun{\agent{j}}{\agent{i}}(\Bfun{\agent{j}}{t{{+}}n{{-}}1} {-} \mx{t{+}n{-}1})\right).$ The rhs equals
        $\mx{t{+}n{-}1} {+} \allowbreak \nicefrac{1}{|\Agents|} \cdot \allowbreak \sum_{\agent{j} {\in} \Agents}  \Ifun{\agent{j}}{\agent{i}}(\Bfun{\agent{j}}{t{+}n{-}1} {-} \mx{t{+}n{-}1})
$.
Let  
$S {=}\sum_{\agent{j} {\in} \Agents \setminus \{\agent{k}\}}   \Ifun{\agent{j}}{\agent{i}}(\Bfun{\agent{j}}{t{{+}}n{{-}}1} {-} \mx{t{+}n{-}1})$ and  
$F {=} \nicefrac{1}{|\Agents|}\Ifun{\agent{k}}{\agent{i}}(\Bfun{\agent{k}}{t{{+}}n{{-}}1} {-} \mx{t{+}n{-}1}).$
 Therefore,
  $\Bfun{\agent{i}}{t{{+}}n} {\leq} \mx{t{+}n{-}1} {+} \nicefrac{1}{|\Agents|}S  {+} F. $   Since $\mx{t{+}n{-}1} \geq \Bfun{\agent{j}}{t{{+}}n{{-}}1}$ for each $j {\in} \Agents$ then  $S {\leq} 0$. We thus obtain $\Bfun{\agent{i}}{t{{+}}n} {\leq} \mx{t{+}n{-}1} {+} \allowbreak F  {=} \allowbreak \mx{t{+}n{-}1} (1 {-} \nicefrac{1}{|\Agents|}\Ifun{\agent{k}}{\agent{i}}) {+} \nicefrac{1}{|\Agents|}\Ifun{\agent{k}}{\agent{i}}\Bfun{\agent{k}}{t{{+}}n{{-}}1}.$  Notice that $\nicefrac{1}{|\Agents|}\Ifun{\agent{k}}{\agent{i}}$ is in $[0,1]$ and from Cor.\ref{cor:mbefore-mafter}, $\mx{t{+}n{-}1} {\leq} \mx{t}$. We can then conclude $\Bfun{\agent{i}}{t{{+}}n}   {\leq}   \mx{t} {+} \nicefrac{1}{|\Agents|}\Ifun{\agent{k}}{\agent{i}} \cdot \allowbreak \left(\Bfun{\agent{k}}{t{{+}}n{{-}}1} {-} \mx{t}\right)$ as wanted. 
 \qed
    \end{enumerate}
\end{proof}

We will use the following notation involving the limits in Cor.\ref{cor:max-limits-exist}. 

\begin{definition}[Min Influence]\label{def:inf-min} Define $\IfunM$ as the smallest positive influence in the influence graph $\Inter$.  Furthermore let 
$L=\lim_{t\to\infty} \mn{t}$ and $U = \lim_{t\to\infty} \mx{t}.$ 
\end{definition}




The path bound lemma states that the belief of agent $i$ at time $t$ is a factor bounding the belief of $j$ at  $t+|p|$  where $p$ is an influence path from $i$ and $j$.  

\begin{lemma}[Path Bound] \label{lemma:path:influence} Assume that  $\Inter$ is strongly connected.
\begin{enumerate}
\item  Let  $p$ be an arbitrary path $\linfl{\agent{i}}{C}{p}{\agent{j}}$.  Then $\Bfun{\agent{j}}{t+|p|} \leq \mx{t} + \nicefrac{C}{|\Agents|^{|p|}}\, (
\Bfun{\agent{i}}{t} - \mx{t}).$
\item Let $\mstar^t \in \Agents$ be a minimal agent at time $t$ and $p$ be a path such that $\linfl{\agent{\mstar}^t}{}{p}{\agent{i}}$. Then
        $\Bfun{\agent{i}}{t{+}|p|} \leq \mx{t} - \delta^t$, with $\delta^t = \left(\nicefrac{\IfunM}{|\Agents|}\right)^{|p|}(U-L)$.
\end{enumerate}

\end{lemma}
\begin{proof} 
\begin{enumerate}
\item Let $p$ be the path $\ldinfl{\agent{i_0}}{C_1}{\agent{i_1}}\ldinfl{}{C_2}{}\ldots \ldinfl{}{C_n} {\agent{i_n}}.$ We show that $\Bfun{\agent{i_n}}{t+|p|} \leq  \mx{t} + \nicefrac{C}{|\Agents|^{|p|}}(
\Bfun{\agent{i_0}}{t} - \mx{t}) $ where $C=C_1\times \ldots \times C_n.$ We proceed by induction on $n$. For $n=1$ we obtain the result immediately from Prop.\ref{prop:upper-bound-inequality}(2).  Assume that $\Bfun{\agent{i_{n-1}}}{t+|p'|} \leq \mx{t} + \nicefrac{C'}{|\Agents|^{|p'|}}(
\Bfun{\agent{i_0}}{t} - \mx{t})$  where $p' = \ldinfl{\agent{i_0}}{C_1}{\agent{i_1}}\ldinfl{}{C_2}{}\ldots \ldinfl{}{C_{n-1}} {\agent{i_{n-1}}}$
and $C' =C_1\times \ldots \times C_{n-1}$  with $n> 1$. Notice that $|p'|=|p|-1$.
Using Prop.\ref{prop:upper-bound-inequality}(2) we obtain
$\Bfun{\agent{i_n}}{t+|p|} \leq \mx{t} + \nicefrac{C_n}{|\Agents|}(\Bfun{\agent{i_{n-1}     }}{t+|p'|} - \mx{t}).$ Using our assumption  
we obtain $\Bfun{\agent{i_n}}{t+|p|} \leq \mx{t} + \nicefrac{C_n}{|\Agents|}( \mx{t} + \nicefrac{C'}{|\Agents|^{|p'|}}(
\Bfun{\agent{i_0}}{t} - \mx{t})  - \mx{t})= \mx{t} + \nicefrac{C}{|\Agents|^{|p|}}(
\Bfun{\agent{i_0}}{t} - \mx{t})$ as wanted.

\item Suppose that $p$ is the path $\linfl{\agent{\mstar}^t}{C}{p}{\agent{i}}$. From Lem.\ref{lemma:path:influence}(1) we obtain  $\Bfun{\agent{i}}{t+|p|} \leq \mx{t} + \nicefrac{C}{|\Agents|^{|p|}}(\Bfun{\agent{\mstar}^t}{t} - \mx{t}) = \mx{t} + \nicefrac{C}{|\Agents|^{|p|}}(\mn{t} - \mx{t}).$  Since $\nicefrac{C}{|\Agents|^{|p|}}(\mn{t} - \mx{t}) \leq 0$, we can substitute $C$ with $\IfunM^{|p|}$. Hence $\Bfun{\agent{i}}{t+|p|} \leq \mx{t} + \left(\nicefrac{\IfunM}{|\Agents|}\right)^{|p|} \cdot (\mn{t}-\mx{t}).$  From Cor.\ref{cor:max-limits-exist}, the maximum value of $\mn{t}$ is $L$ and the minimum value of $\mx{t}$ is $U$, thus $\Bfun{\agent{i}}{t+|p|}{\leq}\mx{t}{+} \left(\nicefrac{\IfunM}{|\Agents|}\right)^{|p|}(L-U){=}\mx{t}{-}\delta^t.$
\qed
\end{enumerate} 
\end{proof}


The following lemma tells us that all the beliefs at time $t+|A|-1$ are smaller than maximal belief at time $t$ by a factor of at least $\epsilon.$

\begin{lemma}[$\epsilon$-Bound]\label{lemma:gamma-bound} Suppose that $\Inter$ is strongly connected.
\begin{enumerate}
  \item If $\Bfun{\agent{i}}{t{+}n} \leq \mx{t} - \gamma$ and $\gamma \geq 0$ then $\Bfun{\agent{i}}{t+n+1} \leq \mx{t} - \nicefrac{\gamma}{|\Agents|}.$
 \item  $\Bfun{\agent{i}}{t+|\Agents|-1} \leq \mx{t} - \epsilon$ where $\epsilon= \left(\nicefrac{\IfunM}{|\Agents|}\right)^{|\Agents|-1}(U-L)$.
\end{enumerate}
\end{lemma}
\begin{proof}  
\begin{enumerate}
 \item Using Prop.\ref{prop:upper-bound-inequality}(1) followed by the assumption $\Bfun{\agent{i}}{t{+}n} \leq \mx{t} - \gamma$ for $\gamma \geq 0$ and the fact that $\Ifun{\agent{j}}{\agent{i}} \in [0,1]$  we obtain the inequality 
$\Bfun{\agent{i}}{t+n+1} \leq   \mx{t} - \gamma + \nicefrac{1}{|\Agents|}\sum_{\agent{j} \in \Agents\setminus \{\agent{i}\}}\Ifun{\agent{j}}{\agent{i}}\left(\Bfun{\agent{j}}{t{+}n} - (\mx{t} - \gamma)\right).$  From Cor.\ref{cor:mbefore-mafter} we have  $\mx{t}\geq \mx{t+n} \geq \Bfun{\agent{j}}{t{+}n} $ for every $\agent{j}\in \Agents$, hence we  obtain  $\Bfun{\agent{i}}{t+n+1} \leq \mx{t} - \gamma + \nicefrac{1}{|\Agents|}\sum_{\agent{j} \in \Agents\setminus \{\agent{i}\}}\Ifun{\agent{j}}{\agent{i}}\left(\mx{t} - (\mx{t} - \gamma)\right).$ Since $\Ifun{\agent{j}}{\agent{i}} \in [0,1]$  we derive $\Bfun{\agent{i}}{t+n+1} \leq \mx{t} - \gamma + \nicefrac{1}{|\Agents|}\sum_{\agent{j} \in \Agents\setminus \{\agent{i}\}}\gamma = \mx{t} - \nicefrac{\gamma}{|\Agents|}.$ 
\item   Let $p$ be the path $\linfl{\agent{\mstar}^t}{}{p}{\agent{i}}$ where $\mstar^t \in \Agents$ is minimal agent at time $t$ and let $\delta^t =  \left(\nicefrac{\IfunM}{|\Agents|}\right)^{|p|}(U-L)$. If $|p| = |\Agents|-1$ then the result follows from Lem.\ref{lemma:path:influence}(2). 
 Else $|p| < |\Agents|-1$ by Def.\ref{def:influence-path}.  We first show by induction on $m$ that $\Bfun{\agent{i}}{t+|p|+m} \leq \mx{t} - \nicefrac{\delta^t}{|\Agents|^m}$ for every $m\geq 0$. If $m=0$, $\Bfun{\agent{i}}{t+|p|} \leq \mx{t} - \delta^t$ by Lem.\ref{lemma:path:influence}(2). 
If $m>0$ and $\Bfun{\agent{i}}{t+|p|+(m-1)} \leq \mx{t} - \nicefrac{\delta^t}{|\Agents|^{m-1}}$ then $\Bfun{\agent{i}}{t+|p|+m} \leq \mx{t} - \nicefrac{\delta^t}{|\Agents|^{m}}$ by Lem.\ref{lemma:gamma-bound}(1).  
Therefore, take $m=|\Agents|-|p|-1$ to obtain  $\Bfun{\agent{i}}{t+|\Agents|-1} \leq \mx{t} - \nicefrac{\delta^t}{|\Agents|^{|\Agents|-|p|-1}}=\mx{t} - \nicefrac{\IfunM^{|p|}.(U-L)}{|\Agents|^{|\Agents|-1}}\leq  \mx{t} - \epsilon$ as wanted.
\qed
\end{enumerate}
\end{proof}

%

Notice that Lem.\ref{lemma:gamma-bound}(2) tells us that $\max^{\cdot}$ decreases by at least $\epsilon$ after  $|A|-1$ steps. Therefore, after $m(|A|-1)$ steps it should decrease by at least $m\epsilon$. 
 
\begin{corollary}\label{cor:max-dif} If $\Inter$ is strongly connected,
    $\mx{t+m(|\Agents|-1)}{\leq}\mx{t}{-}m\epsilon$ for $\epsilon$ in  Lem.\ref{lemma:gamma-bound}(2).
\end{corollary}
%

We can now state that in strongly connected influence graphs extreme beliefs eventually converge to the same value. 

\begin{theorem}[Extreme Consensus] \label{th:U=L} If $\Inter$ is a strongly connected influence graph then   
    $\lim_{t\to\infty} \mx{t} =  \lim_{t\to\infty} \mn{t}.$ 
\end{theorem}

\begin{proof}
Suppose that $\lim_{t\to\infty} \mx{t}=U \neq L=\lim_{t\to\infty} \mn{t}$  for the sake of contradiction.   Let  $\epsilon = \left(\nicefrac{\IfunM}{|\Agents|}\right)^{|\Agents|-1}(U-L).$ From the assumption $U > L$ therefore $\epsilon > 0$. Take $t=0$ and $m=\left(\ceil{\nicefrac{1}{\epsilon}}+1\right)$. Using  
Cor.\ref{cor:max-dif} we obtain  $\mx{0}  \geq \mx{{m(|\Agents|-1)}} + m\epsilon.$ Since $m\epsilon > 1$ and $\mx{m(|\Agents|-1)} \geq 0$ then $\mx{0}> 1.$ But this contradicts Def.\ref{def:extreme:beliefs} which states that $\mx{0} \in [0,1]$.\qed
\end{proof}

Since extreme beliefs converge to the same value, so will the others.  This is an immediate consequence of Th.\ref{th:U=L} and the squeeze theorem \cite{Sohrab:14}.

\begin{theorem}[Belief Consensus]\label{theorem:scc-convergence} Suppose that $\Inter$ is a strongly connected graph. Then for each 
    $\agent{i},\agent{j} \in \Agents, \lim_{t\to\infty} \Bfun{\agent{i}}{t} = \lim_{t\to\infty} \Bfun{\agent{j}}{t}.$
\end{theorem}

The convergence result for strongly connected components implies that for cliques. Recall, however, that not only did we determine belief convergence for cliques but also the convergence value and even a time bound to reach any positive opinion difference (see Lem.\ref{prop:max-min-order-convergence}). Nevertheless, in the next section we consider a meaningful family of strongly connected graphs, which includes cliques and circular graphs, for which we can determine the value of convergence.  

It should be noticed if the influence graph is not strongly connected we cannot guarantee that polarization will disappear; 
all simulations in the fourth row of Fig.\ref{fig:comparing-num-bins} concern a not strongly connected graph (more specifically, a disconnected one), 
and in all of them polarization does not converge to 0.

\subsection{Balanced Influence}

The following notion is inspired by the \emph{circulation problem} for directed graphs (or flow network) \cite{Diestel:17}.  
Given a graph $G=(V,E)$ and a function $c:E \to\reals$ (called \emph{capacity}), the problem involves finding a function  $f:E \to\reals$ (called \emph{flow}) such that  (1) $f(e)\leq c(e)$ for each $e \in E$ and (2) $\sum_{(v,w)\in E}f(v,w)=\sum_{(w,v)\in E}f(w,v)$ for each $v\in V$. If such an $f$ exists it  is called a \emph{circulation} for $G$ and $c$. 
\todo[fancyline]{Frank: I Corrected this definition}

If we think of flow as influence then the second condition, called \emph{flow conservation}, corresponds to requiring that each agent influences the others as much as she is influenced by them; i.e., agent influence should be balanced. 

\begin{definition}[Balanced Influence] \label{def:circulation} We say that  $\Inter$ is \emph{balanced} (or a \emph{circulation}) if every $\agent{i} \in \Agents$ satisfies the constraint $\sum_{\agent{j} \in \Agents} \Ifun{i}{j} = \sum_{\agent{j} \in \Agents} \Ifun{j}{i}.$
\end{definition}

Notice that cliques are balanced. Other examples of balanced influence include circular graphs (see Fig.\ref{fig:interaction-graphs-circular}), where all (non-zero) influence values are equal. More generally, it is easy to see that an influence graph $\Inter$ is balanced if it is a solution to a circulation problem for some $G=(\Agents,\Agents \times \Agents)$ with capacity $c:\Agents\times\Agents \to [0,1].$

\begin{figure}[tb]
    \centering
    \begin{subfigure}{.4\textwidth}
      \centering
      \includegraphics[width=\linewidth]{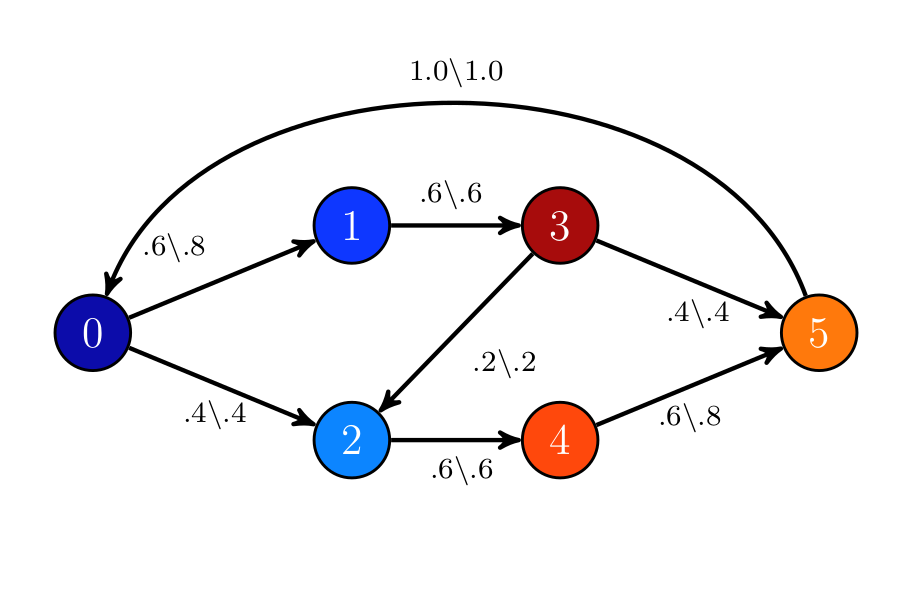}
      \caption{Circulation $\Inter$. Each edge $(i,j)$ in the figure is labeled as $\Ifun{i}{j}\setminus c(i,j)$.}
      \label{fig:circulation-graph}
    \end{subfigure}
    \hfill
    \begin{subfigure}{.5\textwidth}
      \centering
      \includegraphics[width=0.8\linewidth]{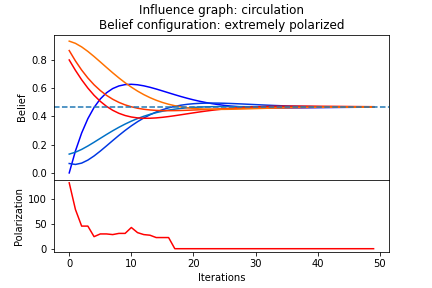}
      \label{fig:circulation-polarization}
      \caption{Evolution of belief and polarization under an extreme polarized initial belief configuration}
    \end{subfigure}
    \caption{Evolution of beliefs and polarization in a circulation graph}
    \label{fig:circulation}
\end{figure}

\begin{example} Fig.\ref{fig:circulation} illustrates a circulation problem  with a balanced influence $\Inter$ as solution. It also shows an evolution of beliefs and polarization under $\Inter$. Notice that the solution is neither  a clique nor a circular graph.
\todo[fancyline]{"Frank:Improve this description"}
\end{example}


An invariant property of balanced influence is that the overall sum of beliefs remains the same across time.

\begin{lemma}\label{lemma:circulation-summation} If  the influence graph $\Inter$ is balanced then $\sum_{\agent{i} \in \Agents} \Bfun{i}{t} = \sum_{\agent{i} \in \Agents} \Bfun{i}{t+1}$.
\end{lemma}

\begin{proof} By Def.\ref{def:rational-update-function} we have
    $\sum_{\agent{i} \in \Agents} \Bfun{i}{t+1} = \sum_{\agent{i} \in \Agents} \left(\Bfun{i}{t} + \nicefrac{1}{|\Agents|} \sum_{\agent{j} \in \Agents} \Ifun{j}{i} (\Bfun{j}{t} - \Bfun{i}{t})\right)$ hence $\sum_{\agent{i} \in \Agents} \Bfun{i}{t+1} = \sum_{\agent{i} \in \Agents} \Bfun{i}{t} + \nicefrac{1}{|\Agents|}\left( \sum_{\agent{i} \in \Agents}\sum_{\agent{j} \in \Agents} \Ifun{j}{i}\Bfun{j}{t} - \sum_{\agent{i} \in \Agents}\sum_{\agent{j} \in \Agents} \Ifun{j}{i}\Bfun{i}{t}\right).$  Using the fact that $\sum_{\agent{i} \in \Agents}\sum_{\agent{j} \in \Agents} \Ifun{j}{i}\Bfun{j}{t}= \sum_{\agent{i} \in \Agents}\sum_{\agent{j} \in \Agents} \Ifun{i}{j}\Bfun{i}{t}$ and Def. \ref{def:circulation}  we obtain $\sum_{\agent{i} \in \Agents} \Bfun{i}{t+1} = \sum_{\agent{i} \in \Agents} \Bfun{i}{t} + \nicefrac{1}{|\Agents|} \sum_{\agent{i} \in \Agents}\sum_{\agent{j} \in \Agents} \Bfun{i}{t}(\Ifun{j}{i}-\Ifun{i}{j})=\sum_{\agent{i} \in \Agents}\Bfun{i}{t}.$ 
\end{proof}

We shall use a fundamental property from flow networks describing flow conservation for graph cuts \cite{Diestel:17}. Interpreted in our case it tells us  that any group of agents $A \subseteq \Agents$  influences other groups as much as they influence  $A$. 

\begin{proposition}[Group Influence Conservation]\label{prop:group-influence-conservation}  Suppose that $\Inter$ is balanced. Let $\{A,B\}$ be a partition of $\Agents$. Then 
$\sum_{i \in A}\sum_{j \in B} \Ifun{i}{j} =\sum_{i \in A}\sum_{j \in B} \Ifun{j}{i}$. 
\end{proposition}
\begin{proof} Immediate consequence of  Prop. 6.1.1 in \cite{Diestel:17}. 
\end{proof}

We now define \emph{weakly connected influence}. Recall that an undirected graph is connected if there is path between each pair of nodes. 

\begin{definition}[Weakly Connected Influence]\label{def:weakly-connected}
Given an influence graph $\Inter$ define the undirected graph $G_{\Inter}=(\Agents,E)$ where $\{i,j \} \in E$ if and only if $\Ifun{i}{j} > 0$ or $\Ifun{j}{i} > 0$.
An influence graph $\Inter$ is called weakly connected if  the undirected graph $G_{\Inter}$ is connected.
\end{definition}

Weakly connected influence is a significant relaxation of its strongly connected counterpart. Nevertheless we shall show next that every weakly connected influence that is balanced it also strongly connected. The flow network intuition behind this property is that a circulation flow never leaves  strongly connected components. 

\begin{lemma}\label{prop:circulation-path}
    If $\Inter$ is balanced and $\Ifun{i}{j} > 0$ then $\Path{j}{i}$.
\end{lemma}
\begin{proof}
    For the sake of contradiction, assume  that $\Inter$ is balanced (a circulation) and $\Ifun{i}{j} > 0$ but there is no path from $\agent{j}$ to $\agent{i}$. Define the agents reachable from $j$, $R_j = \{k \in 
     \Agents | \ \Path{j}{k} \} \cup \{j\}$ and let $\overline{R}_j = \Agents\setminus R_j$. Notice that $(R_j ,\overline{R}_j)$ is a partition of $\Agents.$ Since  the codomain of $\Inter$ is $[0,1]$, $\agent{i} \in \overline{R}_j$, $\agent{j} \in R_j$ and $\Ifun{i}{j} > 0$ we obtain 
     $\sum_{k \in R_j}\sum_{l \in \overline{R}_j} \Ifun{l}{k} > 0$. Clearly there is no $k \in R_j, l \in \overline{R}_j$ such that $\Ifun{k}{l} > 0$, therefore $\sum_{k \in R_j}\sum_{l \in \overline{R}_j} \Ifun{k}{l} = 0$ which contradicts Prop.\ref{prop:group-influence-conservation}. \qed
\end{proof}

\begin{theorem}[Weakly-Connected Balanced Influence]\label{th:balanced-weakly-connected}
    If the influence graph $\Inter$ is balanced and weakly connected then $\Inter$ is also strongly connected. 
\end{theorem}
\begin{proof}
    Immediate consequence of Lem.\ref{prop:circulation-path} and Def. \ref{def:weakly-connected}. 
\end{proof}

It then follows that in a scenario in which every agent influences every other agent directly or indirectly and they influence as much as they are influenced, their beliefs converge to the average of initial beliefs.

\begin{theorem}[Consensus Value]\label{cor:circulation-convergence} If the influence graph $\Inter$ is balanced and weakly connected then  $\lim_{t \to \infty} \Bfun{i}{t} = \nicefrac{1}{|\Agents|}\sum_{\agent{j} \in \Agents}\Bfun{j}{0}.$
\end{theorem}
\begin{proof}
    It is an immediate consequence of Th.\ref{th:balanced-weakly-connected}, Lem. \ref{lemma:circulation-summation} and Th. \ref{theorem:scc-convergence}.
\end{proof}

We conclude this section on balanced influence by highlighting the importance of Th.\ref{th:balanced-weakly-connected}. Recall that $\Inter$ being strongly connected implies belief converge to the same value (Th. \ref{theorem:scc-convergence}) which in turn implies that polarization disappears (i.e., it converges to zero). Then Th.\ref{th:balanced-weakly-connected} tells us that if  polarization does not disappear then either $\Inter$ is not weakly connected or $\Inter$ is not balanced. If $\Inter$ is not weakly connected then there must be isolated subgroups of agents. E.g., consider a scenario with two isolated strongly-connected components; the members of the same component will achieve consensus but the consensus values of the two components may be very different. 
This is illustrated in the fourth row of Fig.~\ref{fig:comparing-num-bins},
representing the simulations for a disconnected influence graph (schematically depicted in Fig.~\ref{fig:interaction-graphs-disconnected}).
Now if $\Inter$ is not balanced then there there must be at least one agent that influences more than what he is influenced (or vice versa). E.g., consider the fifth 
row of Fig.~\ref{fig:comparing-num-bins}, representing the simulations for
a pair of unrelenting influencers (schematically depicted in Fig.~\ref{fig:interaction-graphs-unrelenting}) with high influence over other agents but little or no 
influence from other agents over themselves.

\section{Confirmation Bias}
\label{sec:confirmation-bias}

The previous sections show that the update function of  
Def.~\ref{def:rational-update-function}, although simple,
can shed light on important aspects of polarization.
%
As a step further, here we consider a refinement of the model that 
can capture bias not only towards agents, but towards 
opinions themselves.
This bias, known in social psychology as \emph{confirmation bias}~\cite{Aronson10},
\todo{Sophia, what's the citation you used for confirmation bias in the previous paper?}
is manifest when an agent tends to give more weight to evidence supporting 
their current beliefs than to evidence contradicting them, independently
from its source.

We can incorporate confirmation bias into our
model as follows.
When agent $\agent{j}$ presents agent $\agent{i}$ with evidence for or against proposition $p$, the update function will still move agent $\agent{i}$'s belief toward the belief of agent $\agent{j}$, proportionally to the influence $\Ifun{\agent{j}}{\agent{i}}$ that $\agent{j}$ has over $\agent{i}$, but with a caveat: the move is stronger when $\agent{j}$’s belief is similar to $\agent{i}$’s than when it is dissimilar. 
This is formalized as follows.

\begin{definition}[Confirmation-bias update-function] \label{def:confirmation-bias}
    The \emph{confirmation-bias update-function} $\UpdCB{:}\Blf{\times}\Inter {\rightarrow}\Blf$ 
    is defined analogously to the update function
    of Definition~\ref{def:rational-update-function}, with
    the sole difference that the effect of agent $\agent{j}$ on agent $\agent{i}$'s belief at 
    time $t{+}1$, once both agents interact at time $t$, 
    is given by 
    $\Bapp{i}{j}{t}{=}\Bfun{i}{t}{+}\CBfun{i}{j}{t} \, \Ifun{j}{i} \, (\Bfun{j}{t}{-}\Bfun{i}{t})$,
    where $\CBfun{i}{j}{t}{\in}[0,1]$ is a \emph{confirmation-bias factor} proportional to the difference in the agents beliefs, defined as $\CBfun{i}{j}{t}{=}1{-}|\Bfun{j}{t}{-}\Bfun{i}{t}|$.
\end{definition}

\begin{figure}[tb]
\centering
\begin{subfigure}[t]{0.27\textwidth}
      \centering
      \includegraphics[width=\textwidth]{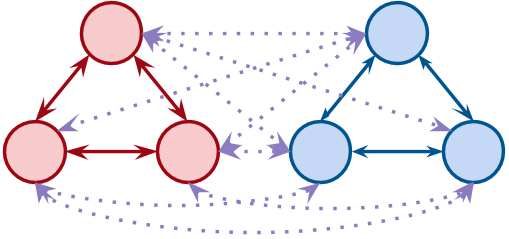}
      \caption{Scheme of faintly communicating influence}
      \label{fig:interaction-graphs-faint}
\end{subfigure}
\hfill
\begin{subfigure}[t]{.36\textwidth}
  \centering
  \includegraphics[width=.8\linewidth]{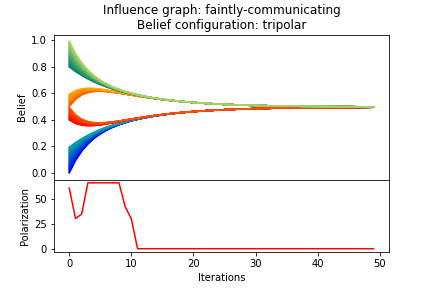}
  \caption{Original update function}
  \label{fig:comparing-classic-update}
\end{subfigure}%
\hfill
\begin{subfigure}[t]{.36\textwidth}
  \centering
  \includegraphics[width=.8\linewidth]{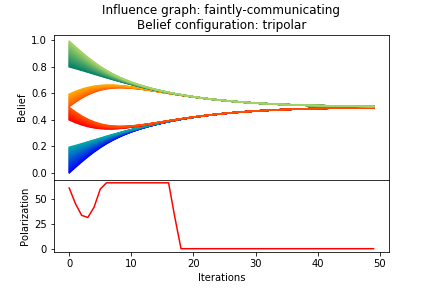}
  \caption{Confirmation-bias update function with
  constant factor $\CBfun{i}{j}{t}{=}0.5$}
  \label{fig:comparing-confirmation-bias}
\end{subfigure}
\caption{Evolution of polarization for tripolar initial belief configuration 
and faintly-communication influence, with $n{=}100$ agents}
\label{fig:comparing-update-functions}
\end{figure}

The incorporation of confirmation bias into our model
allows for the uncovering of new, interesting phenomena.
Fig.~\ref{fig:comparing-update-functions} illustrates
the evolution of polarization, starting from a
tripolar belief configuration, 
in a \emph{faintly communicating} 
influence graph (schematically depicted in Fig.~\ref{fig:interaction-graphs-faint}) 
representing  a social network divided into two groups that 
evolve mostly separately, with little communication between 
them.\footnote{More precisely:
    $\Ifunfaint{\agent{i}}{\agent{j}}{=}0.5$, if $\agent{i},\agent{j}$ are both
    ${\leq} \ceil{\nicefrac{n}{2}}$ or both ${>} \ceil{\nicefrac{n}{2}}$, 
    and $\Ifunfaint{\agent{i}}{\agent{j}}{=}0.1$ otherwise.} 
If agents employ the original update (Fig.\ref{fig:comparing-classic-update}), they are quickly influenced towards a midpoint between their beliefs,
reaching consensus and eliminating polarization.
On the other hand, if agents employ confirmation-bias update
(Fig.\ref{fig:comparing-confirmation-bias}),
convergence happens significantly more slowly, extending
polarization further in time.
Note that, in this case, confirmation bias emphasizes the
non-monotonicity of polarization: since
each group reaches an internal consensus  relatively quickly,
polarization increases; but as each group slowly communicates,
general consensus is eventually achieved.

Note that the confirmation-bias update function generalizes Def.\ref{def:rational-update-function}, since the latter
is the special case of the former in which the confirmation-bias 
factor is always 1.
The proof of Th.\ref{theorem:scc-convergence} can
be easily modified (see Appendix~\ref{sec:proofs}) to show that, 
also under confirmation bias,
in strongly connected influence graphs all agents' beliefs eventually converge to the same value (with the sole exception of a society in which all agents possess extreme beliefs of 0 or 1).
This implies that polarization vanishes over time, even if at a slower pace, in the absence of external influences.
This is formalized below.

\begin{theorem}[Partial belief consensus under confirmation bias]\label{theorem:scc-convergence-confirmation-bias} 
In a strongly connected influence graph and under the confirmation-bias update-function, if there exists an agent $\agent{k} \in \Agents$ such that $\Bfun{k}{0} \notin \{0,1\}$ then for every $\agent{i}, \agent{j} \in \Agents$, $\lim_{t \to \infty} \Bfun{i}{t} = \lim_{t \to \infty} \Bfun{j}{t}$. Otherwise for every $\agent{i} \in \Agents$, $\Bfun{i}{t} = \Bfun{i}{t+1} \in \{0,1\}$.
\end{theorem}


\section{Conclusions}
\label{sec:conclusion}

We considered a model for polarization and belief evolution for multi-agents systems.
We showed that in strongly-connected influence graphs, agents always reach consensus,
so polarization goes to $0$ regardless of initial beliefs or whether agents have confirmation-bias. In the absence of confirmation bias, for balanced families we provided the actual consensus value, and for cliques also the time after which a given opinion difference $\epsilon >0$ is reached. We showed that if polarization does not disappear then either there is an isolated group of agents or there is an agent that influences more than what he is in influenced (Th.\ref{th:balanced-weakly-connected}). We showed  weakly-connected influence graphs where polarization does not converge to zero and several simulations illustrating  polarization in the model. 
We believe that our formal results bring some insights about the phenomenon 
of polarization, which may help in the design of more robust computational models 
for human cognitive and social processes.

As future work we will study belief updates based on the \emph{backfire effect},
which is manifest when, in the face of contradictory evidence, agents' beliefs 
are not corrected, but rather remain unaltered or even get stronger.
The effect has been demonstrated in experiments by psychologists and sociologists~\cite{XXX,YYY,ZZZ}.

\bibliographystyle{splncs04}
\bibliography{polar}

\appendix
\section{Proofs}
\label{sec:proofs}


\subsection{Strongly Connected Influence (Confirmation Bias)}

\begin{remark}All proofs in this section use the \emph{confirmation-bias update-function} from Def. \ref{def:confirmation-bias}, thus note that, although sometimes Theorems look the same as the ones in Section 4, their claims differ in the update-function used, which sometimes we will not highlight for the sake of brevity. Also assume, from now on, that there are no two agents $\agent{i}, \agent{j} \in \Agents$ such that $\Bfun{i}{0} = 0$ and $\Bfun{j}{0} = 1$ thus for every $\agent{i}, \agent{j} \in \Agents$: $\CBfun{\agent{i}}{\agent{j}}{0} > 0$. This might look like a bold assumption but in the end of this section we will address the cases in which this condition does not hold.
\end{remark}
We need to following equation which can be obtained directly from Def.\ref{def:confirmation-bias}. 
\begin{equation}\label{eq:cb-symplification}
    \Bfun{\agent{i}}{t{+}1} =  \Bfun{\agent{i}}{t} + \frac{1}{|\Agents|}\sum_{\agent{j} \in \Agents}\CBfun{i}{j}{t}\Ifun{\agent{j}}{\agent{i}}(\Bfun{\agent{j}}{t} -\Bfun{\agent{i}}{t}) = \Bfun{\agent{i}}{t} + \frac{1}{|\Agents|}\sum_{\agent{j} \in \Agents \setminus \{\agent{i}\}} \CBfun{i}{j}{t}\Ifun{\agent{j}}{\agent{i}}(\Bfun{\agent{j}}{t} -\Bfun{\agent{i}}{t})
\end{equation}

The following lemma states a distinctive property of strongly connected influence: \emph{The belief value of any agent at any time is bounded by those from extreme agents in the previous time unit}. 

\begin{lemma}[Belief Bounds] \label{lemma:cb-maxdiffmin} Assume that the influence graph $\Inter$ is strongly connected. Then  for every $i\in \Agents$,   $\mn{t} \leq \Bfun{\agent{i}}{t{+}1} \leq \mx{t}.$
\end{lemma}
\begin{proof}
We want to prove $\Bfun{\agent{i}}{t{+}1} \leq \mx{t}$. Since $\Bfun{\agent{j}}{t} \leq \mx{t}$, we can use Eq.\ref{eq:symplification} 
to derive the inequality $\Bfun{\agent{i}}{t{+}1} 
    \leq E_1\defsymbol\Bfun{\agent{i}}{t} + \frac{1}{|\Agents|}\sum_{\agent{j} \in \Agents}  \CBfun{\agent{i}}{\agent{j}}{t}\Ifun{\agent{j}}{\agent{i}}(\mx{t} -\Bfun{\agent{i}}{t}).$ Furthermore, $E_1 \leq E_2 \defsymbol\Bfun{\agent{i}}{t} + \frac{1}{|\Agents|}\sum_{\agent{j} \in \Agents} (\mx{t} -\Bfun{\agent{i}}{t})$ because $\CBfun{\agent{i}}{\agent{j}}{t}\Ifun{\agent{j}}{\agent{i}} \leq 1$ and $\mx{t} - \Bfun{\agent{i}}{t} \geq 0.$ We thus obtain $\Bfun{\agent{i}}{t{+}1} \leq E_2 = \Bfun{\agent{i}}{t} + \frac{|\Agents|}{|\Agents|}(\mx{t} -\Bfun{\agent{i}}{t}) = \Bfun{\agent{i}}{t} + \mx{t} -\Bfun{\agent{i}}{t} = \mx{t}$ as wanted. The proof of $\mn{t} \leq \Bfun{\agent{i}}{t{+}1}$ is similar. \qed
\end{proof}

\begin{corollary}
    For all times $t$, there are no agents $\agent{i}, \agent{j} \in \Agents$ such that $\Bfun{i}{t} = 0$ and $\Bfun{j}{t} = 1$, thus for every $\agent{i}, \agent{j} \in \Agents$: $\CBfun{\agent{i}}{\agent{j}}{t} > 0$.
\end{corollary}

\begin{proof}
It is an immediate consequence of the assumption we made that $\agent{i}, \agent{j} \in \Agents$ such that $\Bfun{i}{0} = 0$ and $\Bfun{j}{0} = 1$ and Lemma \ref{lemma:cb-maxdiffmin}
\end{proof}

An immediate consequence of the above lemma, the next corollary tells us that $\mn{\cdot}$ and $\mx{\cdot}$ are monotonically increasing and decreasing functions. 
    
\begin{corollary}[Monotonicity of Extreme Beliefs]\label{cor:cb-mbefore-mafter} If the influence graph $\Inter$ is strongly connected then 
  $\mx{t+1} \leq \mx{t}$ and $min^{t+1} \geq \mn{t}$ for all $t\in\nat$.
  \end{corollary}

The above monotonicity property and the fact that both $\mx{\cdot}$ and $\mn{\cdot}$ are bounded between $0$ and $1$ lead us, via the monotonic convergence theorem \cite{Diestel:17}, to the existence of \emph{limits for the beliefs of extreme agents}.

\begin{corollary}[Limits of Extreme Beliefs]\label{cor:cb-max-limits-exist} If the influence graph $\Inter$ is strongly connected then 
    $\underset{t\to\infty}{\lim} \mx{t} = U$ and $\underset{t\to\infty}{\lim} \mn{t} = L$ for some $U$, $L \in [0,1]$.
\end{corollary}

Another distinctive property of agents under strongly component influence is that the belief of any agent at time $t$ will influence every other agent by the time $t+|\Agents|-1$. The precise statement of this property is given below in the Path Influence Lemma (Lem.\ref{lemma:cb-path:influence}). First we need the following rather technical proposition to prove the lemma.

\begin{proposition}\label{prop:cb-upper-bound-inequality} Assume
$\Inter$ is strongly connected. Let $i,k \in \Agents$, $n,t \in \nat$ with $n\geq 1$, and $v \in [0,1].$
   \begin{enumerate} 
\item  If $\Bfun{\agent{i}}{t} \leq v$ then 
    $\Bfun{\agent{i}}{t{+}1} \leq v + \frac{1}{|\Agents|}\sum_{\agent{j} \in \Agents}\CBfun{i}{j}{t}\Ifun{\agent{j}}{\agent{i}}\left(\Bfun{\agent{j}}{t}-v\right).$ 
 \item $\Bfun{\agent{i}}{t+n} \leq \mx{t} + \frac{1}{|\Agents|}\CBfun{i}{k}{t+n-1}\Ifun{\agent{k}}{\agent{i}}(\Bfun{\agent{k}}{t+n-1} - \mx{t}).$ 
     \end{enumerate}
\end{proposition}
\begin{proof} Let $i,k \in \Agents$, $n,t \in \nat$ with $n\geq 1$, and $v \in [0,1].$
\begin{enumerate}
\item From Def.\ref{def:confirmation-bias}  we obtain $\Bfun{\agent{i}}{t{+}1}  =  \frac{1}{|\Agents|}\sum_{\agent{j} \in \Agents} \left(\Bfun{\agent{i}}{t}(1-\CBfun{i}{j}{t}\Ifun{\agent{j}}{\agent{i}}) + \CBfun{i}{j}{t}\Ifun{\agent{j}}{\agent{i}}\Bfun{\agent{j}}{t}\right)$. If $\Bfun{\agent{i}}{t} \allowbreak \leq v$ then $\Bfun{\agent{i}}{t{+}1} \leq  \frac{1}{|\Agents|}\sum_{\agent{j} \in \Agents} \big(v(1-\CBfun{i}{j}{t}\Ifun{\agent{j}}{\agent{i}}) + \CBfun{i}{j}{t}\Ifun{\agent{j}}{\agent{i}}\Bfun{\agent{j}}{t}\big)$ which is equal to $ v + \frac{1}{|\Agents|}\sum_{\agent{j} \in \Agents} \CBfun{i}{j}{t}\Ifun{\agent{j}}{\agent{i}}\big(\Bfun{\agent{j}}{t} - v\big)$.  

\item By Def.\ref{def:confirmation-bias} $\Bfun{\agent{i}}{t+n} = \frac{1}{|\Agents|}\sum_{\agent{j} \in \Agents}\big(\Bfun{\agent{i}}{t+n-1} + \CBfun{i}{j}{t+n-1}\Ifun{\agent{j}}{\agent{i}}(\Bfun{\agent{j}}{t+n-1} -\Bfun{\agent{i}}{t+n-1})\big).$ From Cor.\ref{cor:cb-mbefore-mafter} we have $\Bfun{\agent{i}}{t+n} \leq \mx{t+n} \leq \mx{t+n-1}$. Thus by Prop.\ref{prop:cb-upper-bound-inequality}(1):  
\begin{align}
        \Bfun{\agent{i}}{t+n} &\leq \frac{1}{|\Agents|}\sum_{\agent{j} \in \Agents}\left(\mx{t+n-1} +   \CBfun{i}{j}{t+n-1}\Ifun{\agent{j}}{\agent{i}}(\Bfun{\agent{j}}{t{+}n{-}1} - \mx{t+n-1})\right) \nonumber \\
        &= \mx{t+n-1} + \frac{1}{|\Agents|}\sum_{\agent{j} \in \Agents}  \CBfun{i}{j}{t+n-1}\Ifun{\agent{j}}{\agent{i}}(\Bfun{\agent{j}}{t+n-1} - \mx{t+n-1}) \nonumber
    \end{align} 
Let us define the sum 
$S =\sum_{\agent{j} \in \Agents \setminus \{\agent{k}\}} \CBfun{i}{j}{t+n-1}\Ifun{\agent{j}}{\agent{i}}(\Bfun{\agent{j}}{t{+}n{-}1} - \mx{t+n-1})$ and the fraction  
$F = \frac{1}{|\Agents|}\CBfun{i}{k}{t+n-1}\Ifun{\agent{k}}{\agent{i}}(\Bfun{\agent{k}}{t{+}n{-}1} - \mx{t+n-1}).$
 Therefore,
  $\Bfun{\agent{i}}{t{+}n} \leq \mx{t+n-1} + \frac{1}{|\Agents|}S  + F. $   Since $\mx{t+n-1} \geq \Bfun{\agent{j}}{t{+}n{-}1}$ for each $j \in \Agents$ then  $S \leq 0$. We thus obtain $\Bfun{\agent{i}}{t{+}n} \leq \mx{t+n-1} + F  = \mx{t+n-1} (1 - \frac{1}{|\Agents|}\CBfun{i}{k}{t+n-1}\Ifun{\agent{k}}{\agent{i}}) + \frac{1}{|\Agents|}\CBfun{i}{k}{t+n-1}\Ifun{\agent{k}}{\agent{i}}\Bfun{\agent{k}}{t{+}n{-}1}.$  Notice that $\frac{1}{|\Agents|}\CBfun{i}{k}{t+n-1}\Ifun{\agent{k}}{\agent{i}}$ is in $[0,1]$ and from Cor.\ref{cor:mbefore-mafter}, $\allowbreak \mx{t+n-1}  \leq \mx{t}$. We can then conclude $\Bfun{\agent{i}}{t{+}n}   \leq   \mx{t} + \frac{1}{|\Agents|}\CBfun{i}{k}{t+n-1}\Ifun{\agent{k}}{\agent{i}}\allowbreak \big(\Bfun{\agent{k}}{t{+}n{-}1} - \mx{t}\big)$ as wanted. 
 
    \end{enumerate}
    \qed
\end{proof}

\begin{definition}[Min Confirmation-bias]\label{def:fcb-min}
Denote by $\CBfunM = \min_{\agent{i}, \agent{j} \in \Agents} \CBfun{i}{j}{0}$ the minimum confirmation-bias factor in our graph throughout time. \todo[fancyline]{Frank: The explanation shouldn't be part of the definition: "Since  $\mn{t}$ and $\mx{t}$ ... I moved it outside}
\end{definition}
Notice that since $\mn{t}$ and $\mx{t}$ get closer as $t$ (Cor.\ref{cor:cb-mbefore-mafter}), $\CBfunM$ does not decrease when time passes, thus, it acts as a lower bound for the confirmation-bias factor in every time step.

The following property intuitively tells us that an agent $i$ at time $t$ will bound the belief of $j$ at time  $t+t'$  where $t'$ is the size of any influence path from $i$ and $j$. The bound depends on the belief of agent $i$ at time $t$ and the product influence $C$ of $i$ over $j$ along the corresponding path.

\begin{lemma}[Path Influence] \label{lemma:cb-path:influence} Assume that  $\Inter$ is strongly connected. Let  $p$ be an arbitrary path $\linfl{\agent{i}}{C}{p}{\agent{j}}$.  Then $\Bfun{\agent{j}}{t+|p|} \leq \mx{t} + \frac{C\CBfunM^{|p|}}{|\Agents|^{|p|}}(
\Bfun{\agent{i}}{t} - \mx{t}).$
\end{lemma}
\begin{proof} Let $p$ be the path $\ldinfl{\agent{i_0}}{C_1}{\agent{i_1}}\ldinfl{}{C_2}{}\ldots \ldinfl{}{C_n} {\agent{i_n}}.$ We show that $\Bfun{\agent{i_n}}{t+|p|} \leq  \mx{t} + \frac{C\CBfunM}{|\Agents|}(
\Bfun{\agent{i_0}}{t} - \mx{t}) $ where $C=C_1\times \ldots \times C_n.$ We proceed by induction on $n$. For $n=1$, since $\Bfun{\agent{i_0}}{t} - \mx{t} \leq 0$  we obtain the result immediately from Prop.\ref{prop:cb-upper-bound-inequality}(2).  Assume that $\Bfun{\agent{i_{n-1}}}{t+|p'|} \leq \mx{t} + \frac{C'\CBfunM^{|p'|}}{|\Agents|^{|p'|}}(
\Bfun{\agent{i_0}}{t} - \mx{t})$  where $p' = \ldinfl{\agent{i_0}}{C_1}{\agent{i_1}}\ldinfl{}{C_2}{}\ldots \ldinfl{}{C_{n-1}} {\agent{i_{n-1}}}$
and $C' =C_1\times \ldots \times C_{n-1}$  with $n> 1$. Notice that $|p'|=|p|-1$.
Using Prop.\ref{prop:cb-upper-bound-inequality}(2) and the fact that $\Bfun{\agent{i_{n-1}}}{t+|p'|} - \mx{t} \leq 0$ we obtain
$\Bfun{\agent{i_n}}{t+|p|} \leq \mx{t} + \frac{C_n\CBfunM}{|\Agents|}(\Bfun{\agent{i_{n-1}     }}{t+|p'|} - \mx{t}).$ Using our assumption  
we obtain $\Bfun{\agent{i_n}}{t+|p|} \leq \mx{t} + \frac{C_n\CBfunM}{|\Agents|}( \mx{t} + \frac{C'\CBfunM^{|p'|}}{|\Agents|^{|p'|}}(
\Bfun{\agent{i_0}}{t} - \mx{t})  - \mx{t})= \mx{t} + \frac{C\CBfunM^{|p|}}{|\Agents|^{|p|}}(
\Bfun{\agent{i_0}}{t} - \mx{t})$ as wanted.
\end{proof}

\begin{theorem}\label{theorem:cb-max-bel-geq-delta-path} Assume that $\Inter$ is strongly connected. Let $\mstar^t \in \Agents$ be a minimal agent at time $t$ and $p$ be a path such that $\linfl{\agent{\mstar}^t}{}{p}{\agent{i}}$. Then
    \begin{center}
        $\Bfun{\agent{i}}{t{+}|p|} \leq \mx{t} - \delta^t$, with $\delta^t = \left(\frac{\IfunM\CBfunM}{|\Agents|}\right)^{|p|}(U-L)$.
    \end{center}
\end{theorem}

\begin{proof} Suppose that $p$ is the path $\linfl{\agent{\mstar}^t}{C}{p}{\agent{i}}$. From Lem.\ref{lemma:cb-path:influence} we obtain  $\Bfun{\agent{i}}{t+|p|} \leq \mx{t} + \frac{C\CBfunM^{|p|}}{|\Agents|^{|p|}}(\Bfun{\agent{\mstar}^t}{t} - \mx{t}) = \mx{t} + \frac{C\CBfunM^{|p|}}{|\Agents|^{|p|}}(\mn{t} - \mx{t}).$  Since $\frac{C\CBfunM^{|p|}}{|\Agents|^{|p|}}(\mn{t} - \mx{t}) \leq 0$, we can substitute $C$ with $\IfunM^{|p|}$. Thus, $\Bfun{\agent{i}}{t+|p|} \leq \mx{t} + \Big(\frac{\IfunM\CBfunM}{|\Agents|}\Big)^{|p|} \allowbreak (\mn{t}-\mx{t}).$  From Cor.\ref{cor:cb-max-limits-exist}, the maximum value of $\mn{t}$ is $L$ and the minimum value of $\mx{t}$ is $U$, thus $\Bfun{\agent{i}}{t+|p|} \leq \mx{t} + \Big(\frac{\IfunM\CBfunM}{|\Agents|}\Big)^{|p|}(L-U) = \mx{t} - \delta^t.$ 
\end{proof}

\begin{proposition}\label{prop:cb-gamma-bound} Suppose that $\Inter$ is strongly connected.
    If $\Bfun{\agent{i}}{t{+}n} \leq \mx{t} - \gamma$ and $\gamma \geq 0$ then $\Bfun{\agent{i}}{t+n+1} \leq \mx{t} - \frac{\gamma}{|\Agents|}.$
\end{proposition}
\begin{proof} Using Prop.\ref{prop:cb-upper-bound-inequality}(1) followed by the assumption $\Bfun{\agent{i}}{t{+}n} \leq \mx{t} - \gamma$ for $\gamma \geq 0$ and the fact that $\Ifun{\agent{j}}{\agent{i}} \in [0,1]$  we obtain the inequality 
$\Bfun{\agent{i}}{t+n+1} \leq   \mx{t} - \gamma + \frac{1}{|\Agents|}\sum_{\agent{j} \in \Agents\setminus \{\agent{i}\}}\CBfun{i}{j}{t+n}\Ifun{\agent{j}}{\agent{i}}\left(\Bfun{\agent{j}}{t{+}n} - (\mx{t} - \gamma)\right).$  From Cor.\ref{cor:mbefore-mafter}  $\mx{t}\geq \mx{t+n} \geq \Bfun{\agent{j}}{t+n} $ for every $\agent{j}\in \Agents$, hence $\Bfun{\agent{i}}{t+n+1} \leq \mx{t} - \gamma + \frac{1}{|\Agents|}\sum_{\agent{j} \in \Agents\setminus \{\agent{i}\}}\CBfun{\agent{i}}{\agent{j}}{t+n}\Ifun{\agent{j}}{\agent{i}}\big(\mx{t} - (\mx{t} - \gamma)\big).$ Since $\CBfun{i}{j}{t+n}\Ifun{\agent{j}}{\agent{i}} \in [0,1]$  we derive $\Bfun{\agent{i}}{t+n+1} \leq \mx{t} - \gamma + \frac{1}{|\Agents|}\sum_{\agent{j} \in \Agents\setminus \{\agent{i}\}}\gamma \allowbreak = \mx{t} - \frac{\gamma}{|\Agents|}.$ \qed
\end{proof}

\begin{theorem}\label{theorem:cb-max-bel-geq-epsilon} If $\Inter$ is strongly connected then 
   $\Bfun{\agent{i}}{t+|\Agents|-1} \leq \mx{t} - \epsilon$, where $\epsilon$ is equal to $\left(\frac{\IfunM\CBfunM}{|\Agents|}\right)^{|\Agents|-1}(U-L)$.
\end{theorem}

\begin{proof}
  Let $p$ be the path $\linfl{\agent{\mstar}^t}{}{p}{\agent{i}}$ where $\mstar^t \in \Agents$ is minimal agent at time $t$ and let $\delta^t =  \left(\frac{\IfunM\CBfunM}{|\Agents|}\right)^{|p|}(U-L)$. If $|p| = |\Agents|-1$ then the result follows from Th.\ref{theorem:cb-max-bel-geq-delta-path}. 
 Else $|p| < |\Agents|-1$ by Def.\ref{def:influence-path}.  We first show by induction on $m$ that $\Bfun{\agent{i}}{t+|p|+m} \leq \mx{t} - \frac{\delta^t}{|\Agents|^m}$ for every $m\geq 0$. If $m=0$, $\Bfun{\agent{i}}{t+|p|} \leq \mx{t} - \delta^t$ by Th.\ref{theorem:cb-max-bel-geq-delta-path}. 
If $m>0$ and $\Bfun{\agent{i}}{t+|p|+(m-1)} \leq \mx{t} - \frac{\delta^t}{|\Agents|^{m-1}}$ then $\Bfun{\agent{i}}{t+|p|+m} \leq \mx{t} - \frac{\delta^t}{|\Agents|^{m}}$ by Prop.\ref{prop:cb-gamma-bound}.  
Therefore, take $m=|\Agents|-|p|-1$ to obtain  $\Bfun{\agent{i}}{t+|\Agents|-1} \leq \mx{t} - \frac{\delta^t}{|\Agents|^{|\Agents|-|p|-1}}=\mx{t} - \frac{(\IfunM\CBfunM)^{|p|}.(U-L)}{|\Agents|^{|\Agents|-1}}\leq  \mx{t} - \epsilon$ as wanted. \qed

\end{proof}

\begin{corollary}
    $\mx{t+m(|\Agents|-1)} \leq \mx{t} - m\epsilon$ for $\epsilon$ as in  Th.\ref{theorem:cb-max-bel-geq-epsilon}.
\end{corollary}

\begin{theorem} 
    If $\Inter$ is strongly connected  then 
    $\lim_{t\to\infty} \mx{t} =  \lim_{t\to\infty} \mn{t}.$ 
\end{theorem}

\begin{proof}
Suppose, by contradiction, that $\lim_{t\to\infty} \mx{t}=U \neq L=\lim_{t\to\infty} \mn{t}$.   Let  $\epsilon = \left(\frac{\IfunM\CBfunM}{|\Agents|}\right)^{|\Agents|-1}(U-L).$ From the assumption $U > L$ therefore $\epsilon > 0$. Take $t=0$ and $m=\left(\ceil{\frac{1}{\epsilon}}+1\right)$. Using  
Cor.\ref{cor:max-dif} we obtain  $\mx{0}  \geq \mx{{m(|\Agents|-1)}} + m\epsilon.$ Since $m\epsilon > 1$ and $\mx{m(|\Agents|-1)} \geq 0$ then $\mx{0}> 1.$ But this contradicts Def.\ref{def:extreme:beliefs} which states that $\mx{0} \in [0,1]$.\qed
\end{proof}

As an immediate consequence of Th.\ref{th:U=L} and the squeeze theorem, we conclude that for strongly-connected graphs, the belief of all agents converge to the same value.

\begin{theorem}\label{theorem:cb-scc-convergence} If $\Inter$ is strongly connected then for each 
    $\agent{i},\agent{j} \in \Agents, \underset{t \to \infty}{\lim} \Bfun{\agent{i}}{t} = \underset{t \to \infty}{\lim} \Bfun{\agent{j}}{t}.$
\end{theorem}
We assumed, in the beginning of this section, that there were no two agents $\agent{i},\agent{j}$ such that $\Bfun{\agent{i}}{t} = 0$ and $\Bfun{j}{t} = 1$, now we will address the situation in which this does not happen, thus, assuming that those agents $i$ and $j$ existed, we are left with two cases. Before showing them, we must prove a technical proposition.

\begin{proposition} [Influencing the extremes] \label{prop:cb-influencing-extremes}
Supposing that $\Inter$ is strongly connected and that there exists an agent $\agent{k}$ such that $\Bfun{k}{0} \notin \{0,1\}$, $\mx{|\Agents-1|} < 1$.
\end{proposition}

\begin{proof}
It is enough to show that, for every $\agent{i}$, being $p$ a path $\linfl{\agent{k}}{}{p}{\agent{i}}$, $\Bfun{i}{|p|} < 1$ such that $p = ki_1i_2\ldots i_{n-1}i_n..i$. We will prove it by induction in $n = |\linfl{\agent{k}}{}{p}{\agent{i_n}}|$.
For $n = 0$, it is true via the hypothesis. For $n > 1$, we have, by the inductive hypothesis and Def. \ref{def:influence-path}, that $\Bfun{i_{n-1}}{|p|-1} < 1$ and $\Ifun{i_{n-1}}{n} > 0$. Thus, $\Bfun{i_n}{|p|} = \Bfun{i_n}{|p|-1} + \frac{1}{|\Agents|}\sum_{\agent{j} \in \Agents} \CBfun{i_n}{j}{|p|-1}\Ifun{j}{i_n} (\Bfun{j}{|p|-1} - \Bfun{i_n}{|p|-1})$ separating $i_{n-1}$ from the summation we get that it $\Bfun{i_n}{|p|} = \Bfun{i_n}{|p|-1} + \frac{1}{|\Agents|}\sum_{\agent{j} \in \Agents\setminus{}\{i_{n-1}\}} \CBfun{i_n}{j}{|p|-1}\Ifun{j}{i_n} (\Bfun{j}{|p|-1} - \Bfun{i_n}{|p|-1}) + \frac{1}{|\Agents|} \CBfun{i_n}{i_{n-1}}{|p|-1}\allowbreak\Ifun{i_{n-1}}{i_n} (\Bfun{i_{n-1}}{|p|-1} - \Bfun{i_n}{|p|-1}) \allowbreak \leq 1 + \frac{1}{|\Agents|}\CBfun{i_n}{i_{n-1}}{|p|-1}\Ifun{i_{n-1}}{i_n} (\Bfun{i_{n-1}}{|p|-1} - 1) < 1$.
\qed
\end{proof}

The following Theorem is the confirmation-bias version of Thr. \ref{theorem:scc-convergence} and tells us that beliefs always converge and that, unless the society is composed only by very extreme agents, they converge to the same value. 
\begin{theorem}[Confirmation-bias belief convergence]\label{theorem:cb-geberal-scc-convergence}
In a strongly connected influence graph and under the confirmation-bias update-function, if there exists an agent $\agent{k} \in \Agents$ such that $\Bfun{k}{0} \notin \{0,1\}$ then for every $\agent{i}, \agent{j} \in \Agents$, $\lim_{t \to \infty} \Bfun{i}{t} = \lim_{t \to \infty} \Bfun{j}{t}$. Otherwise for every $\agent{i} \in \Agents$, $\Bfun{i}{t} = \Bfun{i}{t+1} \in \{0,1\}$.
\end{theorem}

\begin{proof}
\begin{enumerate}
    \item If there exists an agent $\agent{k} \in \Agents$ such that $\Bfun{k}{0} \notin \{0,1\}$, we can use Prop. \ref{prop:cb-influencing-extremes} to show that by the time $|\Agents|-1$ no agent has belief $1$, thus we fall on the general case stated in the beginning of the section (starting at a different time step does not make any difference for this purposes) and, thus, all beliefs converge to the same value according to Thr. \ref{theorem:cb-scc-convergence}.
    \item Otherwise it is easy to see that beliefs remain constant as $0$ or $1$ throughout time, since the agents are so biased towards each other the only agents $\agent{j}$ able to influence another agent $\agent{i}$ ($\CBfun{i}{j}{t} \neq 0$) have the same belief as $\agent{i}$.
\end{enumerate}
\qed
\end{proof}

\section{Axioms for Esteban-Ray polarization measure}
\label{sec:polar-axioms}

The Esteban-Ray polarization measure used in this paper was developed as the only function satisfying all of the following conditions and axioms~\cite{Esteban:94:Econometrica}:
\begin{description}
   \item[Condition H:] 
    The ranking induced by the polarization measure over two
    distributions is invariant w.r.t. the size of the population.
    (This is why we can assume w.l.o.g. that the distribution is a probability distribution.)
    $$
     \PfunER{\pi, y} \geq \PfunER{\pi', y'} \implies \forall \lambda > 0, \PfunER{\lambda\pi, y} \geq \PfunER{\lambda\pi', y'}~.
    $$
    \item[Axiom 1:] 
    Consider three levels of belief $p, q, r \in [0,1]$ 
    such that the same proportion of the population holds beliefs $q$ and $r$, and
    a significantly higher proportion of the population holds belief $p$.
    If the groups of agents that hold beliefs $q$ and $r$ reach a consensus and
    agree on an \qm{average} belief $\nicefrac{(q+r)}{2}$, then the social network becomes
    more polarized.
    \item[Axiom 2:] 
    Consider three levels of belief $p, q, r \in [0,1]$,
    such that $q$ is at least as close to $r$ as it is to $p$, and
    $p > r$.
    If only small variations on $q$ are permitted, the direction that brings it closer 
    to the nearer and smaller opinion ($r$) should increase polarization.
    \item[Axiom 3:] 
    Consider three levels of belief $p, q, r \in [0,1]$, s.t. $p < q < r$ and there 
    is a non-zero proportion of the population holding belief $q$.
    If the proportion of the population that holds belief $q$ is equally split into
    holding beliefs $q$ and $r$, then polarization increases.
\end{description}



\end{document}